\PassOptionsToPackage{noend}{algpseudocode}
\documentclass[sigconf]{acmart}

\copyrightyear{2025}
\acmYear{2025}
\setcopyright{cc}
\setcctype{by-nc-nd}
\acmConference[RecSys '25]{Proceedings of the Nineteenth ACM Conference on Recommender Systems}{September 22--26, 2025}{Prague, Czech Republic}
\acmBooktitle{Proceedings of the Nineteenth ACM Conference on Recommender Systems (RecSys '25), September 22--26, 2025, Prague, Czech Republic}\acmDOI{10.1145/3705328.3748084}
\acmISBN{979-8-4007-1364-4/2025/09}

\usepackage[utf8]{inputenc}
\usepackage{url}
\usepackage{subfig}

\usepackage{researchpack}
\algrenewcommand{\algorithmicrequire}{\textbf{Precondition:}}
\algrenewcommand{\algorithmicensure}{\textbf{Output:}}

\begin{document}

\title[Affect-aware CDR for AT via Music Preference Elicitation]{
  Affect-aware Cross-Domain Recommendation for Art Therapy\texorpdfstring{\\}{ } 
  via Music Preference Elicitation
}

\author{Bereket A. Yilma}
\author{Luis A. Leiva}
\email{name.surname@uni.lu}
\affiliation{%
  \institution{University of Luxembourg}
  \country{Luxembourg}
}

\begin{abstract}
Art Therapy (AT) is an established practice that facilitates emotional processing and recovery through creative expression. Recently, Visual Art Recommender Systems (VA RecSys) have emerged to support AT, demonstrating their potential by personalizing therapeutic artwork recommendations. Nonetheless, current VA RecSys rely on visual stimuli for user modeling, limiting their ability to capture the full spectrum of emotional responses during preference elicitation. Previous studies have shown that music stimuli elicit unique affective reflections, presenting an opportunity for cross-domain recommendation (CDR) to enhance personalization in AT. Since CDR has not yet been explored in this context, we propose a family of CDR methods for AT based on music-driven preference elicitation. A large-scale study with 200 users demonstrates the efficacy of music-driven preference elicitation, outperforming the classic visual-only elicitation approach. Our source code, data, and models are available at \url{https://github.com/ArtAICare/Affect-aware-CDR}.
\end{abstract}



\maketitle

\section{Introduction}
\label{sec:intro}

Art therapy (AT) has long been recognized as a powerful intervention for psychological distress,
including trauma, anxiety, and depression~\cite{malchiodi2011handbook}. 
It provides individuals with a medium for creative expression 
that facilitates emotional processing and self-exploration, fostering recovery and well-being. 
Particularly, a visual exposure form of AT includes a preference elicitation phase, 
where patients select paintings from a curated set of visual stimuli that resonate with their healing journey, 
followed by a guided session featuring a similar set of artworks recommended by art therapists, 
designed to evoke positive emotions and contain healing elements~\cite{haeyen2021imagery}.  

A significant challenge in AT lies in personalizing these interventions 
for individuals with unique traumatic experiences. 
Traditional AT requires the expertise of trained therapists in selecting artworks 
that are therapeutically appropriate for each patient's healing journey. 
This process, while effective, is resource-intensive and constrained by the therapist's familiarity 
with available artworks and the diversity and uniqueness of patients’ emotional needs. 
The vastness of the search space for relevant visual stimuli 
makes it difficult to identify the right pieces for each patient, 
posing a substantial barrier to scalable, personalized interventions~\cite{hathorn2008guide}. 

Recent advancements in digital AT have aimed to improve 
both the accessibility and the effectiveness of therapeutic interventions through computational tools. 
In particular, Visual Art Recommender Systems (VA RecSys) 
have recently been developed to personalize visual stimuli 
based on the psychological states of individuals, showing promising results. 
For example, Yilma et al.~\cite{yilma2024artful, yilma2025ai} 
developed VA RecSys that promote healing and recovery
in the context of Post-Intensive Care Syndrome (PICS) intervention.
While useful, they only considered visual stimuli for preference elicitation. 
However, a different modality, music, has the potential to further elicit affective cues 
that can be missed in visual stimuli~\cite{juslin2008emotional,feng2025effect}. 
Music plays a significant role in eliciting emotional responses~\cite{thaut2013rhythm, ren2024affective}, 
which are essential for psychological healing,
potentially enhancing the personalization and effectiveness of digital AT.

State-of-the-art (SOTA) approaches in digital AT 
have not yet explored the cross-domain recommendation (CDR) potential 
of affective stimuli knowledge transfer between VA and music.  
Hence, we propose a novel approach 
that integrates music as a preference elicitation modality for personalized VA RecSys in AT. 
Concretely, we propose a family of CDR methods:
(1)~\textit{Mozart}, or \textit{affect-aware contrastive alignment of music and paintings},
which maps the affective properties of music and paintings into a shared latent space
that allows a semantic retrieval of therapeutic content across the two modalities;
(2)~\textit{Haydn}, or \textit{affective space search}, based on $k$-nearest neighbors,
which leverages the valence and arousal labels of each modality to construct a similarity matrix for retrieval;
(3)~\textit{Salieri}, or \textit{multimodal alignment of music and paintings}, 
using large language models (LLMs) and vision-language models (VLMs),
which enables the extraction of semantic representations from both modalities, 
facilitating cross-modal mapping for retrieval. 
Our main research question is: \emph{Can music-driven preferences provide therapeutic art recommendations?}

We assess the efficacy of these CDR strategies for AT on a large-scale user study 
against the uni-modal SOTA VA RecSys engine reported in previous work, 
comprising a ResNet-50 backbone and visual-only preference elicitation. 
In sum, we make the following contributions:
\begin{itemize}
    \item We introduce CDR to the field of AT by integrating music and VA domains.
    
    \item We develop and evaluate three novel affect-aware CDR algorithms based on music preferences.
    
    \item We conduct a large-scale user study with 200 people with psychiatric sequelae,
    comparing the proposed approaches against the SOTA baseline.
    
    \item We provide insights on integrating CDR into digital AT 
    and discuss future research for mental health interventions.
\end{itemize}

\section{Background and Related Work}
\label{sec:rl}

\subsection{VA RecSys in Art Therapy}

VA pieces like paintings have been widely utilized 
to promote psychological well-being in clinical environments~\cite{hu2021art}. 
As trauma can often be hard to verbalize in spoken psychotherapy sessions, 
art opens a creative medium of expression~\cite{di2022psychotherapy},
stimulating emotions and self-reflection, 
and addressing various psychological disorders 
such as depression and anxiety~\cite{haeyen2021imagery, stuckey2010connection}.

To offer an effective AT based on the unique needs of patients, 
personalization remains vital yet largely constrained 
due to its reliance on human expertise and a manual art selection process~\cite{abbing2018effectiveness}. 
Very recently, VA RecSys have emerged to support the delivery of personalized AT.
Yilma et al.~\cite{yilma2024artful} developed VA RecSys that leverage visual 
as well as textual features of paintings for semantic retrieval of therapeutic art recommendations. 
The proposed models used Residual Neural Network (ResNet50)~\cite{he2016deep} for image feature extraction, 
Latent Dirichlet Allocation (LDA)~\cite{blei2003latent} 
and Bidirectional Encoder Representations from Transformers (BERT)~\cite{devlin2018bert} 
for textual feature extraction, 
and Bootstrapping Language-Image Pre-training (BLIP)~\cite{li2022blip} 
for multimodal feature extraction.
A follow-up study~\cite{yilma2025ai} investigated a human-in-the-loop approach 
to safeguard recommendations and foster human-AI collaboration in AT.  
While these systems have shown promise, they rely solely on visual inputs, 
potentially missing broader affective dimensions. 
On the other hand, music serves as a complementary modality for preference elicitation, 
owing to its capacity to elicit emotional responses~\cite{juslin2008emotional}. 
Several studies have also linked music to therapeutic neurological responses, 
suggesting its role in preference detection.\cite{koelsch2014brain, vik2019neuroplastic, brancatisano2020music}.

Jazi et al.~\cite{yousefian2021emotion} proposed an affect-aware music RecSys 
that maps user interactions to emotional states, 
while Patel et al.~\cite{patel2024emotion} and Iordanis~\cite{iordanis2021emotion} 
developed Deep Learning models for personalized music recommendations based on affective cues. 
These works highlight music’s potential to reveal affective preferences 
that could inform visual art recommendations. 
However, music as a preference elicitation tool for VA RecSys remains unexplored, 
motivating our investigation into a CDR approach in AT 
via affective knowledge transfer between music and paintings.

\subsection{CDR and the Music-Art Gap}

Cross-domain recommendation leverages user preferences or item features 
from one domain to enhance recommendations in another domain, 
mitigating issues like data sparsity and cold-start problems~\cite{elkahky2015multi, zang2022survey}. 
CDR has been particularly useful in the context of movie recommendations,
using for example book~\cite{zhu2021cross} and music~\cite{fernandez2016alleviating} preferences.
The key idea is to find a suitable embedding and mapping framework~\cite{man2017cross, zhang2021deep, liu2022human}.
Recent works have also explored graph-based methods for CDR~\cite{hirakawa2021cross}

Despite these advances, CDR between music and VA remains underexplored, 
and in particular in therapeutic contexts. 
A study by Campos-Bueno et al.~\cite{campos2015emotional} 
investigated the affective dimensions of valence and arousal in music and VA. 
They found that music elicited stronger effects on valence than paintings. 
It has been reported that both modalities elicit profound emotional responses, 
music through auditory stimulation~\cite{thaut2013rhythm} and VA via imagery~\cite{haeyen2021imagery}, 
yet no framework transfers affective knowledge between them for CDR.
Lee et al.~\cite{lee2020crossing} investigated music-to-art style transfer using generative adversarial networks, 
but their focus was aesthetic rather than therapeutic. 

Emotional responses from music have long been leveraged in music therapy for emotional regulation, 
where specific musical elements such as tempo, melody, and harmony 
are known to influence affective states~\cite{juslin2008emotional, ren2024affective}. 
Recent advancements have applied machine learning techniques to predict the effectiveness of music therapy interventions~\cite{RAGLIO2020105160, jiao2025advancing}. 
Similarly, paintings have been shown to evoke strong emotional reactions, 
with different artistic styles, color palettes, and subject matter being used in AT 
to help individuals process trauma, anxiety, and depression~\cite{haeyen2021imagery, malchiodi2011handbook}. 
Further research in neuroscience underscores the role of affective control in emotion regulation, 
which can support therapeutic outcomes~\cite{schweizer2020role, ladas2024harmony}. 
Recent studies found that active and passive engagement in creative arts 
activates neural circuits that promote adaptive emotional regulation~\cite{barnett2024arts}. 
Scoping reviews of integrated arts therapies and neuroscience 
emphasize the need for interdisciplinary approaches to compare modalities like music and visual arts~\cite{bokoch2025scoping}.

In sum, it is evident that music has potential to evoke profound affective responses.
Hence, our main hypothesis is that the integration of affective cues 
across these two domains can derive better therapeutic content recommendations.
Unfortunately, no affect-aware CDR framework has been proposed to date in a unified therapeutic context. 
To address this critical gap, we propose a family of CDR approaches 
that leverage music clips for preference elicitation to derive therapeutic art recommendations. 

\section{Method: Therapeutic VA RecSys from Music}
\label{sec:ControlRec}

In this work, we depart from utilizing visual stimuli for preference elicitation. 
Instead, we leverage music stimuli to elicit affective preferences for therapeutic painting recommendations. 
To the best of our knowledge, no prior work in digital AT involves cross-domain affective knowledge transfer. 
Thus, we take the first step to introduce affect-aware CDR in digital AT.

\begin{figure*}[!ht]
\centering
\includegraphics[width=\linewidth]{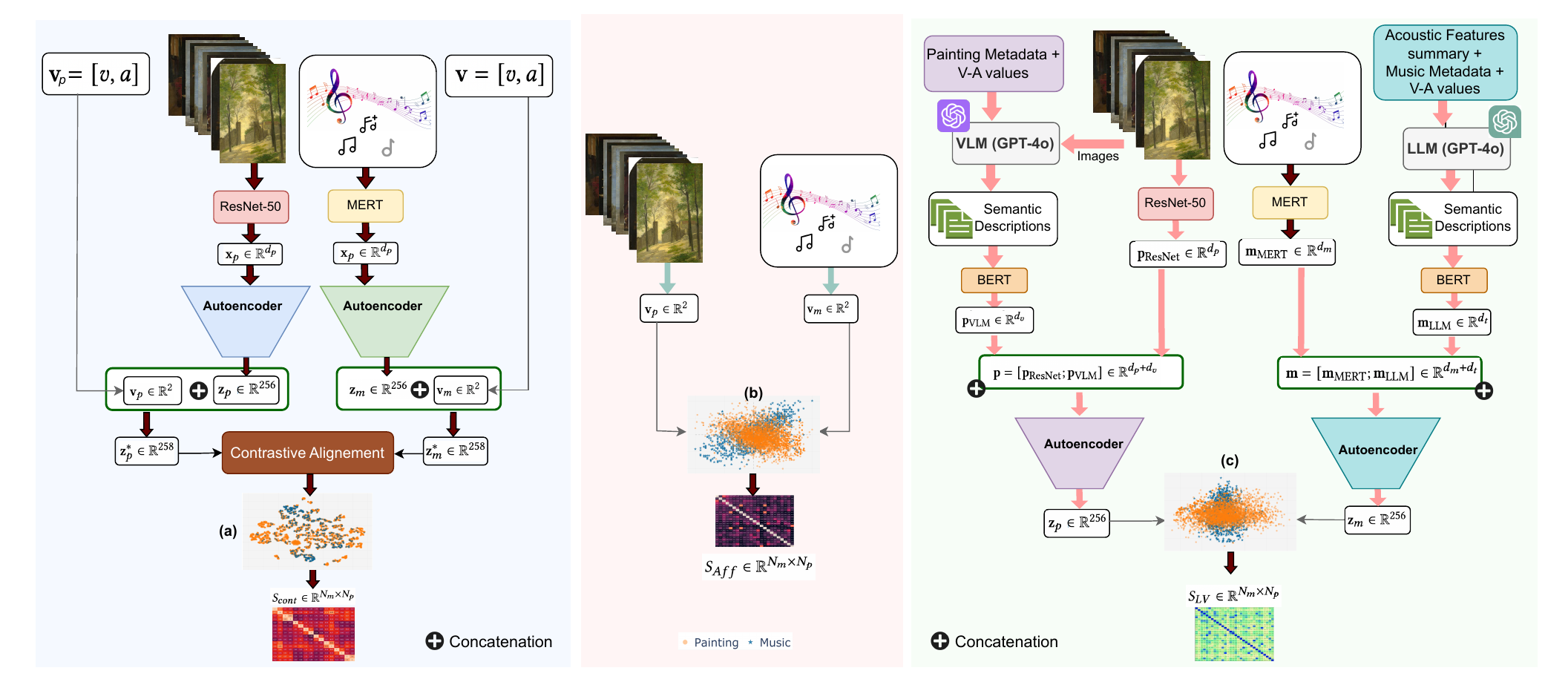}
\caption{
    Proposed system architectures. From left to right: 
    Affect-aware contrastive alignment (Mozart), 
    Affective Space Search (Haydn), 
    and multimodal alignment with LLM and VLM (Salieri).
    Notice that each approach produces different latent embedding spaces (a,b,c).
}
\Description{
 Illustration of the three proposed cross-domain recommendation system architectures arranged from left to right. The left panel depicts the Mozart approach, which employs affect-aware contrastive alignment by extracting features from music using MERT and from paintings using ResNet50, reducing dimensions with autoencoders, concatenating with valence-arousal (V-A) vectors, and applying contrastive learning with a Gaussian kernel-based similarity measure on continuous V-A distances. The middle panel shows the Haydn method, utilizing k-nearest neighbors (kNN) for affective space search directly on V-A vectors to compute Euclidean distances and generate recommendations via weighted averages. The right panel illustrates the Salieri framework, involving multi-modal alignment with large language models (LLMs) for music text generation from acoustic features and V-A prompts, vision-language models (VLMs) for painting descriptions, and similarity computation in a joint embedding space.
}
\label{fig:cross-D}
\end{figure*}

\subsection{Dataset}

Although evidence from both music and AT literature highlights the potential for affect-aware CDR in AT, 
a significant challenge remains in the availability of parallel music and VA datasets with affective annotations.  These datasets need to be carefully curated to ensure they match therapeutic goals 
and reflect the emotional qualities of each modality. 
While there are several datasets for music and paintings individually, 
few are integrated with affective labels, making it difficult to leverage them for the affect-aware CDR task. 
To overcome this limitation, we conducted a thorough analysis 
and selected reliable public datasets from both domains that are well suited for this task. 
In the following, we outline the dataset selection, preprocessing steps, and the proposed CDR strategies.

\subsubsection{Music Dataset}

We selected the DEAM (Database for Emotional Analysis of Music) dataset,\footnote{\url{https://cvml.unige.ch/databases/DEAM/manual.pdf}}
as it has been reliably used in music therapy~\cite{choi2024accelerated,panwar2019you}. 
DEAM contains 1,802 instrumental music excerpts annotated with valence and arousal values both continuously (per-second) and over the whole song. For our purposes, we used the averaged per-song annotations but filtered out songs with high standard deviation to enforce stability and reduce ambiguity. We excluded songs with valence SD > $1.75$ and arousal SD > $1.0$,
chosen based on the label distributions. This filtering resulted in 909 songs.

\subsubsection{Visual Art Dataset}

For paintings, we chose the WikiArt Emotions Dataset,\footnote{\url{https://saifmohammad.com/WebPages/wikiartemotions.html}}
as it is the only public dataset with reliable affective annotations. 
It contains 4,105 paintings annotated based on combined image and metadata analysis. 
Each painting is labeled with multiple emotions and their intensities. 

\subsubsection{Preprocessing}

Both datasets have different labels to represent their emotional triggers on users. 
DEAM is annotated with continuous valence-arousal (V-A) scores in the $[-1,1]$ range, 
while WikiArt is labeled with multiple emotions and their intensities. 
First, we need to convert the labels of the two datasets to uniform affective annotations (i.e., V-A values). 
For DEAM, we considered per-song static annotations 
and filtered out songs with high standard deviation, to ensure stability and reduce ambiguity. 
Then, to align the WikiArt dataset emotional labels with the V-A annotations, 
we first normalized the intensity of each emotion label for each painting
and then we used the `NRC Valence, Arousal, and Dominance' (NRC VAD) Lexicon,
which provides V-A scores for over 20,000 English words,\footnote{Despite some cultural differences, it has been shown that the majority of affective norms are stable across languages~\cite{mohammad2025nrc}.}
to translate these emotions into V-A scores, in the $[-1,1]$ range.

After this, we obtained two modalities, paintings and music; 
both with respective modality-specific features and V-A values. 
The final dataset is comprised of 909 music samples and 4,105 painting samples. 

Although these affective labels are considered to be representative, 
we hypothesize that the feature-level nuances of each modality can play a significant role 
in the emotional reflection triggered by each modality~\cite{chowdhury-2019-towards, aslan2022recognizing}. 
Therefore, we aim to learn a joint representation space 
guided by emotional awareness to derive better recommendations.

\subsection{Mozart: Cross-Domain Recommendation via Affect-Aware Contrastive Alignment}
\label{subsec:cont}

To establish a unified representation space, 
we start by extracting modality-specific features using pre-trained networks: 
MERT~\cite{li2023mert} for music and ResNet50~\cite{he2016deep} for paintings. 
Let $\mathbf{x}_m \in \mathbb{R}^{d_m}$ and $\mathbf{x}_p \in \mathbb{R}^{d_p}$ 
be the extracted feature vectors for music and paintings, respectively, 
where $d_m$ and $d_p$ denote the dimensionality of the raw feature representations from each modality. 
We then obtain reduced representations through modality-specific autoencoders:
\begin{equation}
\mathbf{z}_m = AE_m(\mathbf{x}_m), \quad \mathbf{z}_p = AE_p(\mathbf{x}_p),
\end{equation}
where $AE_m$ and $AE_p$ are autoencoders 
trained with reconstruction loss separately for music and paintings, respectively, 
to learn compact 256D embeddings: $\mathbb{R}^{256}$. 
This ensures that the embeddings retain meaningful information 
while reducing dimensionality.\footnote{256D was chosen to balance expressiveness and efficiency. Higher dimensions increase model complexity and computational cost, while lower dimensions may not sufficiently capture feature-rich representations. 
An offline evaluation (ranking overlap) showed no significant gains with 512D or 1024D.} 
Thus, producing embeddings $\mathbf{z}_m, \mathbf{z}_p \in \mathbb{R}^{256}$.

Since emotional perception is fundamental to both music and VA, 
we explicitly incorporate V-A representations. 
Each sample is associated with a V-A vector $\mathbf{v}_i = [v_i, a_i]$, 
where $v_i$ and $a_i$ denote the valence and arousal values, respectively. 
Finally we concatenate the raw V-A values with the learned embeddings:
\begin{equation}
\mathbf{z}_{m}^* = [\mathbf{z}_m, \mathbf{v}_a], \quad \mathbf{z}_{p}^* = [\mathbf{z}_p, \mathbf{v}_a].
\end{equation}
This results in enriched embeddings $\mathbf{z}_{m}^*, \mathbf{z}_{p}^* \in \mathbb{R}^{258}$ 
that encode both modality-specific features and affective information, 
forming the basis for the subsequent contrastive learning objective.

\subsubsection{Contrastive Learning with Continuous V-A Values}

Traditional contrastive learning frameworks rely on binary labels 
to indicate whether sample pairs are similar or dissimilar. 
In our work, each sample of music and painting is associated with continuous V-A values. 
To effectively capture the degree of emotional similarity, 
we first compute the Euclidean distance\footnote{Unlike cosine similarity, which ignores magnitude, Euclidean distance ensures that emotionally distinct but directionally similar items are properly differentiated.} between the V-A vectors of two samples:
\[
d_{ij} = \|\mathbf{v}_i - \mathbf{v}_j\|_2 = \sqrt{(v_i - v_j)^2 + (a_i - a_j)^2},
\label{eq_dist}
\]
where $\mathbf{v}_i = [v_i, a_i]$ represents the V-A vector for sample $i$. 
To convert this distance into a soft, continuous similarity measure, we apply a Gaussian kernel:
\[
S_{ij} = \exp\left(-\frac{d_{ij}^2}{2\sigma^2}\right),
\label{eq_GausKernel}
\]
where $\sigma$ is a bandwidth parameter that controls how quickly similarity decays with increasing distance.  
This approach is particularly effective for continuous labels, 
as it allows the model to weigh the contribution of each pair according to the degree of emotional closeness. 
Recent work by Tan et al.~\cite{tan2024contrastive} demonstrates that kernel-based similarity measures 
enhances the learning of continuous representations in cross-modal settings. 
In our work, the continuous similarity score $S_{ij}$ is integrated into the contrastive loss:
\[
\mathcal{L} = \sum_{i,j} S_{ij} \left( \|\mathbf{z}_i^* - \mathbf{z}_j^*\|_2 \right)^2 + 
    (1 - S_{ij}) \max\left( 0, m - \|\mathbf{z}_i^* - \mathbf{z}_j^*\|_2 \right)^2,
\label{eq_Loss}
\]
where $\mathbf{z}_i^*$ and $\mathbf{z}_j^*$ are the enriched joint embeddings 
incorporating both modality-specific features and V-A information. 
The margin $m$ ensures a minimum separation between dissimilar samples. 

\subsubsection{Addressing Dataset Imbalance}
\label{sec:weighted-loss}

The disparity in dataset sizes 909 music samples versus 4,105 art samples can bias the learning process. 
To mitigate this imbalance, we adopted a weighted loss approach.
We adjust \eqref{eq_Loss} by assigning weights to each sample based on its modality. 
Define the weighting factors as:
\[
\lambda_m = \frac{N_p}{N_m + N_p}, \quad \lambda_p = \frac{N_m}{N_m + N_p},
\]
where $N_m = 909$ and $N_p = 4105$. 
The weighted contrastive loss then becomes
$\mathcal{L}_{\text{weighted}} = \sum_{i,j} \lambda_i \lambda_j \mathcal{L}$,

where $\lambda_i$ equals $\lambda_m$ if sample $i$ is from the music domain and $\lambda_p$ if from the art domain.

\subsubsection{Implementation details}
Music features were extracted using MERT-v1-330M.\footnote{\url{https://huggingface.co/m-a-p/MERT-v1-330M}}
Audio files were resampled to 24 kHz and converted to mono.
The MERT-based embeddings were concatenated with precomputed acoustic features 
(means, medians, and standard deviations of tempo, spectral contrast, etc.) from DEAM. 
For paintings, a pre-trained ResNet50 model extracted 2048D features 
from images resized to 224×224 pixels, using ImageNet-standard normalization, 
sourced from the WikiArt dataset.
Music and painting embeddings were further reduced to 256D 
using modality-specific autoencoders with encoder layers \texttt{[input\_dim, 1024, 512, 256]} and mirrored decoders, trained with Adam optimizer ($\eta = 10^{-3}, \beta_1=0.9, \beta_2=0.999$) and MSE loss. 
The resulting 256D embeddings were Min-Max scaled to $[-1,1]$
and concatenated with the V-A labels from each dataset,
producing 258D enriched embeddings ($\mathbf{z}_m^*, \mathbf{z}_p^*$) per modality.
Then, for contrastive alignment, a projection head, 
implemented as a two-layer MLP with architecture [258, 256, 128], 
mapped the 258D embeddings to a 128D joint space. 

The modality-specific weights of the weighted contrastive loss 
are $\lambda_m = 0.818$ and $\lambda_p = 0.182$, based on the weighting factors in \autoref{sec:weighted-loss}.
The optimal hyperparameters for the Gaussian kernel-based similarity measure (\autoref{eq_GausKernel}),
i.e., the bandwidth parameter $\sigma$ and the margin $m$, were determined via a grid search. 
The best performance was achieved with $\sigma = 0.5$ and $m = 0.5$.
Mozart was trained for up to 50 epochs using Adam optimizer ($\eta = 10^{-3}, \beta_1=0.9, \beta_2=0.999$)
and early stopping of 5 epochs, monitoring the validation loss.
The resulting joint embedding space, visualized in  \autoref{fig:cross-D}, 
Before, the similarity matrix shows a rather effective alignment, where both modalities contribute to shaping the learned representation 
guided by a combination of affective and modality-specific features.

By computing Euclidean distances between the 128D joint embeddings, 
we get a similarity matrix $\mathbb{R}^{N_m \times N_p}$ between all music and painting pairs, 
enabling efficient retrieval of painting recommendations for music tracks. 
Then, given a user’s list of preferred music tracks \( m_1, m_2, \dots, m_k \) 
with corresponding ratings \( r_1, r_2, \dots, r_k \) on a 5-point Likert scale, 
we normalize the ratings to obtain weights \( w_1, w_2, \dots, w_k \), 
where each \( w_i \in [0, 1] \) and \( \sum_{i=1}^{k} w_i = 1 \). 
The weighted average distance from the set of music tracks to each painting \( p_j \) is then calculated as:
\[
d_{\text{Mozart}}(p_j) = \sum_{i=1}^{k} w_i \cdot \|\mathbf{z}_{m_i}^+ - \mathbf{z}_{p_j}^+\|_2
\label{eq:mozart}
\]
where \(\mathbf{z}_{m_i}^+\) and \(\mathbf{z}_{p_j}^+\) are the 128D joint embeddings. 
The top-\( n \) paintings with the smallest \( d_{\text{Mozart}}(p_j) \) values are recommended, 
reflecting their emotional similarity to the user’s music preferences, weighted by the normalized ratings.

\subsection{Haydn: Cross-Domain Recommendation via Affective Space Search}

We implemented a $k$-nearest neighbors (kNN) VA RecSys approach using affective space search 
which allows retrieval based on the affective alignment of music and painting V-A vectors. 
Each music track \( m_i \) and painting \( p_j \) 
is represented by its V-A vectors \(\mathbf{v}_{m_i}\) and \(\mathbf{v}_{p_j}\), respectively. 
The affective distance between them is computed using the Euclidean distance, 
as defined in \autoref{eq_dist},
yielding a similarity matrix $\mathbb{R}^{N_m \times N_p}$ 
between all music and painting pairs.

Similar to Mozart, a weighted average is computed to make recommendations 
from a precomputed similarity matrix derived from Euclidean distances, 
with weights \( w_i \in [0, 1] \) and \( \sum_{i=1}^{k} w_i = 1 \).  
The weighted average affective distance 
from the set of music tracks to each painting \( p_j \) is calculated as:
\[
d_{\text{Haydn}}(p_j) = \sum_{i=1}^{k} w_i \cdot \|\mathbf{v}_{m_i} - \mathbf{v}_{p_j}\|_2
\label{eq:haydn}
\]
where \(\mathbf{v}_{m_i}\) and \(\mathbf{v}_{p_j}\) are the V-A vectors of music $m_i$ and painting $p_j$ respectively. This is a straightforward and interpretable method, though it may not fully capture modality-specific attributes responsible for the different emotional reflections triggered on users.

\subsection{Salieri: Cross-Domain Recommendation via multimodal Alignment with LLM and VLM}

We propose a third approach based on large language models (LLMs) and vision-language models (VLMs), as these models demonstrate a strong capability in capturing subtle emotional and semantic details, which are crucial for affective cross-modal retrieval. Salieri provides a complementary perspective to both the affective space search and contrastive alignment approaches discussed earlier. This approach constructs multimodal representations for both music and paintings by leveraging transformer-based models for feature extraction and semantic embedding generation. The overall pipeline consists of three main stages: (i)~modality-specific feature extraction, (ii)~multimodal embedding construction, and (iii)~cross-modal similarity computation.

\subsubsection{Music Representation Learning}
For music, we extract two distinct feature representations. 
First, we learn acoustic features using MERT similar to the contrastive approach in Mozart (\autoref{subsec:cont}) 
to obtain deep audio embeddings, denoted as $\mathbf{m}_{\text{MERT}} \in \mathbb{R}^{d_m}$. 
Next, we learn textual representations of music's semantic description by querying an LLM (GPT-4o) 
with metadata including the V-A values.
The resulting textual descriptions were then converted to an embedding 
using the transformer-based language model BERT, yielding $\mathbf{m}_{\text{LLM}} \in \mathbb{R}^{d_t}$.

The final music embedding is obtained by concatenating these two representations:
\begin{equation}
    \mathbf{m} = [\mathbf{m}_{\text{MERT}} ; \mathbf{m}_{\text{LLM}}] \in \mathbb{R}^{d_m + d_t}
\end{equation}
To ensure a compact representation, 
we train an autoencoder $f_m: \mathbb{R}^{d_m + d_t} \to \mathbb{R}^{256}$ 
that reduces the dimensionality of $\mathbf{m}$ to $\mathbf{z}_m \in \mathbb{R}^{256}$.

\subsubsection{Painting Representation Learning}
For paintings, a similar two-stream feature extraction strategy is employed. 
Similar to Mozart, we extracted visual features of paintings using ResNet-50,
obtaining $\mathbf{p}_{\text{ResNet}} \in \mathbb{R}^{d_p}$. 
Subsequently for textual representation, we queried a VLM model (GPT-4o) with the painting images, their metadata, valence-arousal values, and emotion labels from WikiArt. 
The textual descriptions from GPT-4o  were then converted to an embedding 
using the transformer-based language model BERT to produce $\mathbf{p}_{\text{VLM}} \in \mathbb{R}^{d_v}$.

The overall painting embedding is constructed by concatenation:
\begin{equation}
    \mathbf{p} = [\mathbf{p}_{\text{ResNet}} ; \mathbf{p}_{\text{VLM}}] \in \mathbb{R}^{d_p + d_v}
\end{equation}
As with music, we employ an autoencoder $f_p: \mathbb{R}^{d_p + d_v} \to \mathbb{R}^{256}$ 
to obtain a reduced representation $\mathbf{z}_p \in \mathbb{R}^{256}$.

\subsubsection{Cross-Modal Similarity Computation}
Given the 256D embeddings $\mathbf{z}_m$ and $\mathbf{z}_p$ for music and paintings, respectively, 
we compute the similarity matrix $S_{LV} \in \mathbb{R}^{N_m \times N_p}$ 
between all music and painting pairs using cosine similarity. 
As in Mozart, a weighted average is computed to make recommendations 
based on user preference ratings using this similarity matrix:
\[
d_{\text{Salieri}}(p_j) = \sum_{i=1}^{k} w_i \cdot \|\mathbf{z}_{m_i} - \mathbf{z}_{p_j}\|_2
\label{eq:salieri}
\]
This approach allows us to compare LLM and VLM-based representations 
with both the affective space search and contrastive alignment techniques in CDR for AT. 

\begin{algorithm}[!ht]
\small
\caption{CDR preprocessing pipeline}
\label{algo:preprocessing}
\begin{algorithmic}[1]
    \Procedure{PreprocessCDR}{$M, P$}
        \State $x_m \gets$ \Call{ExtractMERT}{$M$} \Comment{Mozart music features}
        \State $x_p \gets$ \Call{ExtractResNet50}{$P$} \Comment{Mozart VA features}
        \State $\mathbf{v}_{m}, \mathbf{v}_{p} \gets$ \Call{GetVAScores}{$M, P$} \Comment{Haydn}
        \State $z_m^* \gets$ Reduce ($x_m$, 256D) + $\mathbf{v}_{m}$ $\rightarrow \mathbb{R}^{258}$ \Comment{Mozart}
        \State $z_p^* \gets$ Reduce ($x_p$, 256D) + $\mathbf{v}_{p}$ $\rightarrow \mathbb{R}^{258}$ \Comment{Mozart}
        \State $z_m^+, z_p^+ \gets$ Reduce ($z_m^*, z_p^*$, 128D) \Comment{Mozart}
        \State $m_{\text{MERT}} \gets$ \Call{ExtractMERT}{$M$} \Comment{Salieri music features}
        \State $m_{\text{LM}} \gets$ \Call{ExtractLLM}{$M$} \Comment{Salieri music text features}
        \State $p_{\text{ResNet}} \gets$ \Call{ExtractResNet50}{$P$} \Comment{Salieri VA features}
        \State $p_{\text{VLM}} \gets$ \Call{ExtractVLM}{$P$} \Comment{Salieri VA text features}
        \State $m \gets$ [$m_{\text{MERT}}; m_{\text{LM}}$] \Comment{Salieri concatenation}
        \State $p \gets$ [$p_{\text{ResNet}}; p_{\text{VLM}}$] \Comment{Salieri concatenation}
        \State $z_m, z_p \gets$ Reduce ($m, p$, 256D) \Comment{Salieri}
        \State \Return $\mathbf{v}_{m}, \mathbf{v}_{p}, z_m^+, z_p^+, z_m, z_p, p_{\text{ResNet}}$
    \EndProcedure
\end{algorithmic}
\end{algorithm}

\begin{algorithm}[!ht]
\small
\caption{CDR recommendations}
\label{algo:recommendation}
\begin{algorithmic}[1]
    \Require Extracted features for each engine (\autoref{algo:preprocessing}).
    \Procedure{RecommendCDR}{$M, P, R$}
        \State $L \gets d(p_j)$ using ratings $w_i \in R$ \Comment{See \eqref{eq:mozart} \eqref{eq:haydn} \eqref{eq:salieri}}
        \State \Call{Sort}{$L$} \Comment{Ascending order}
        \State \Return \Call{Slice}{$L, n$} \Comment{Top-$n$ paintings}
    \EndProcedure
\end{algorithmic}
\end{algorithm}

\begin{figure*}[!ht]
    \centering
    \def\w{0.19\linewidth}
    \subfloat[Pre-study questions\label{fig:web:before}]{
        \includegraphics[width=\w]{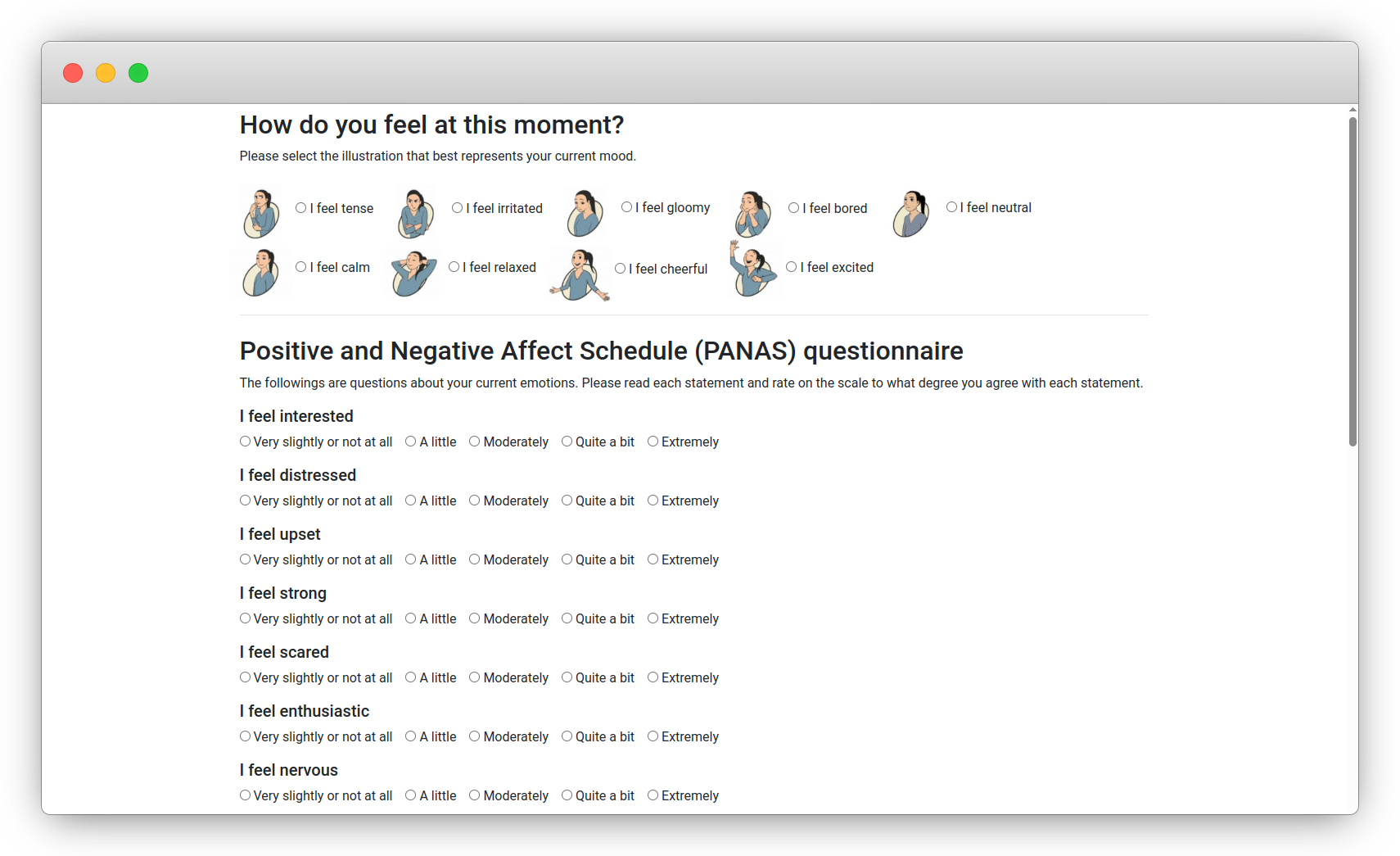}
    }
    \subfloat[Preference elicitation\label{fig:web:elicitation}]{
        \includegraphics[width=\w]{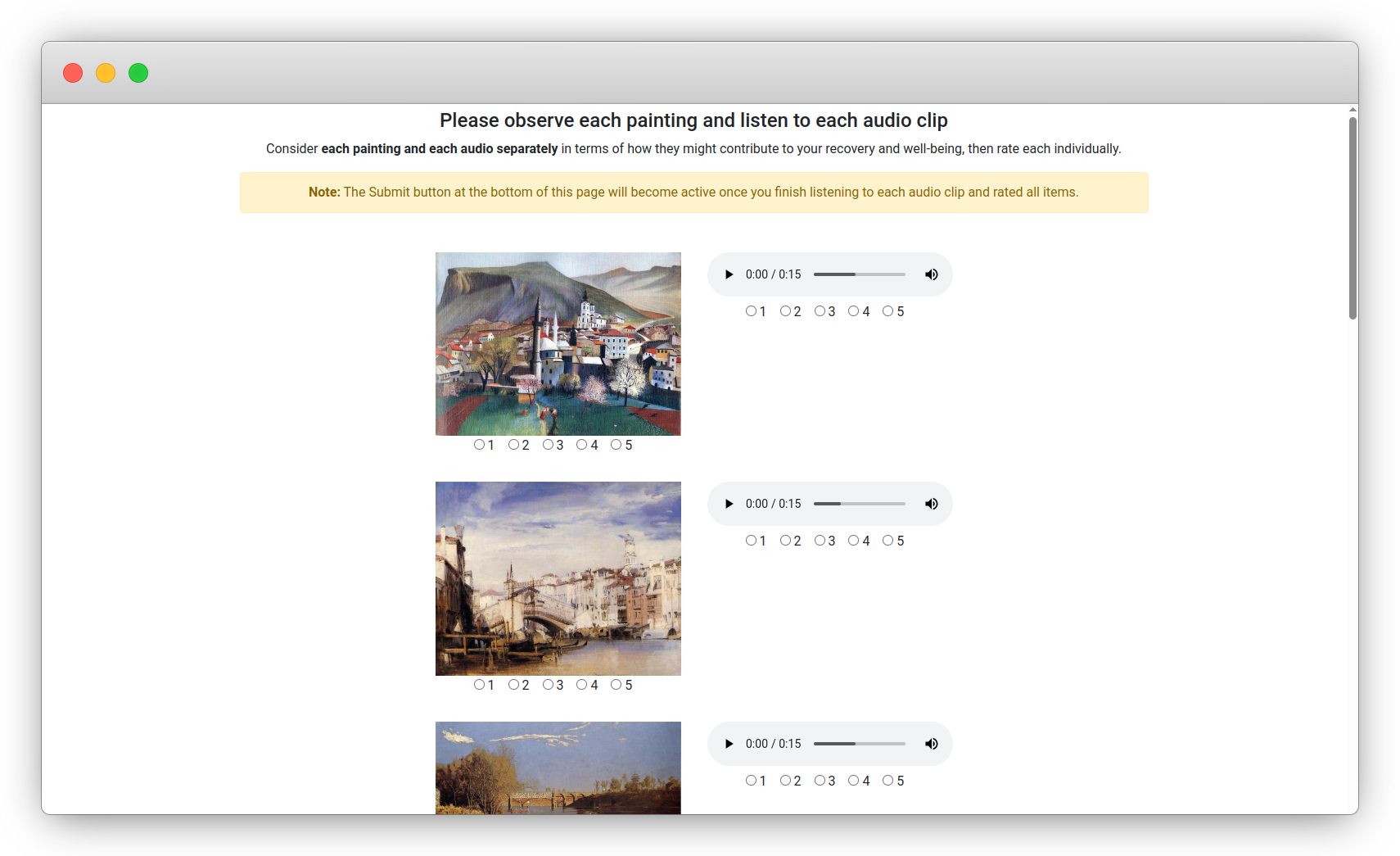}
    }
    \subfloat[Guided AT session\label{fig:web:ratings}]{
        \includegraphics[width=\w]{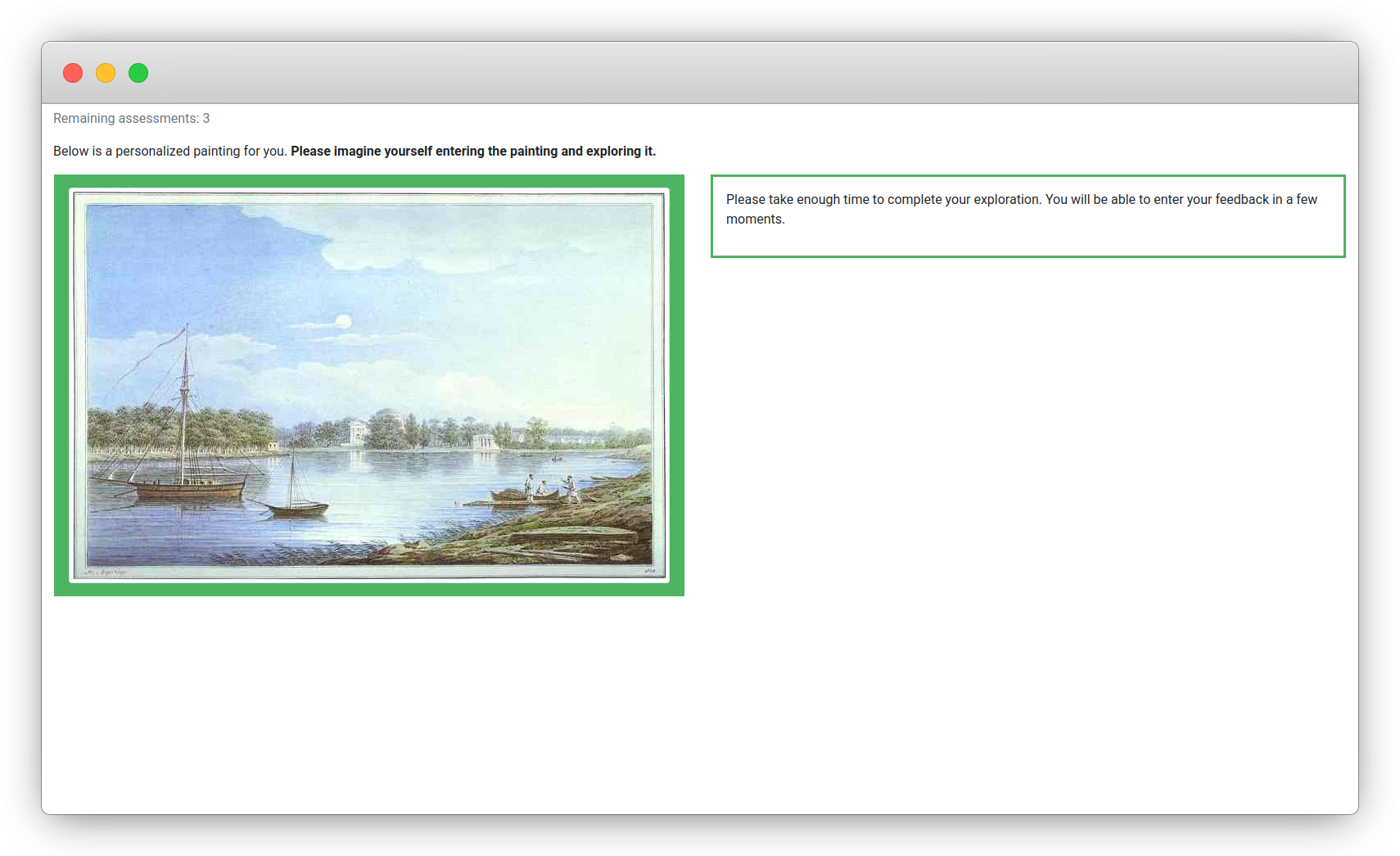}
        \includegraphics[width=\w]{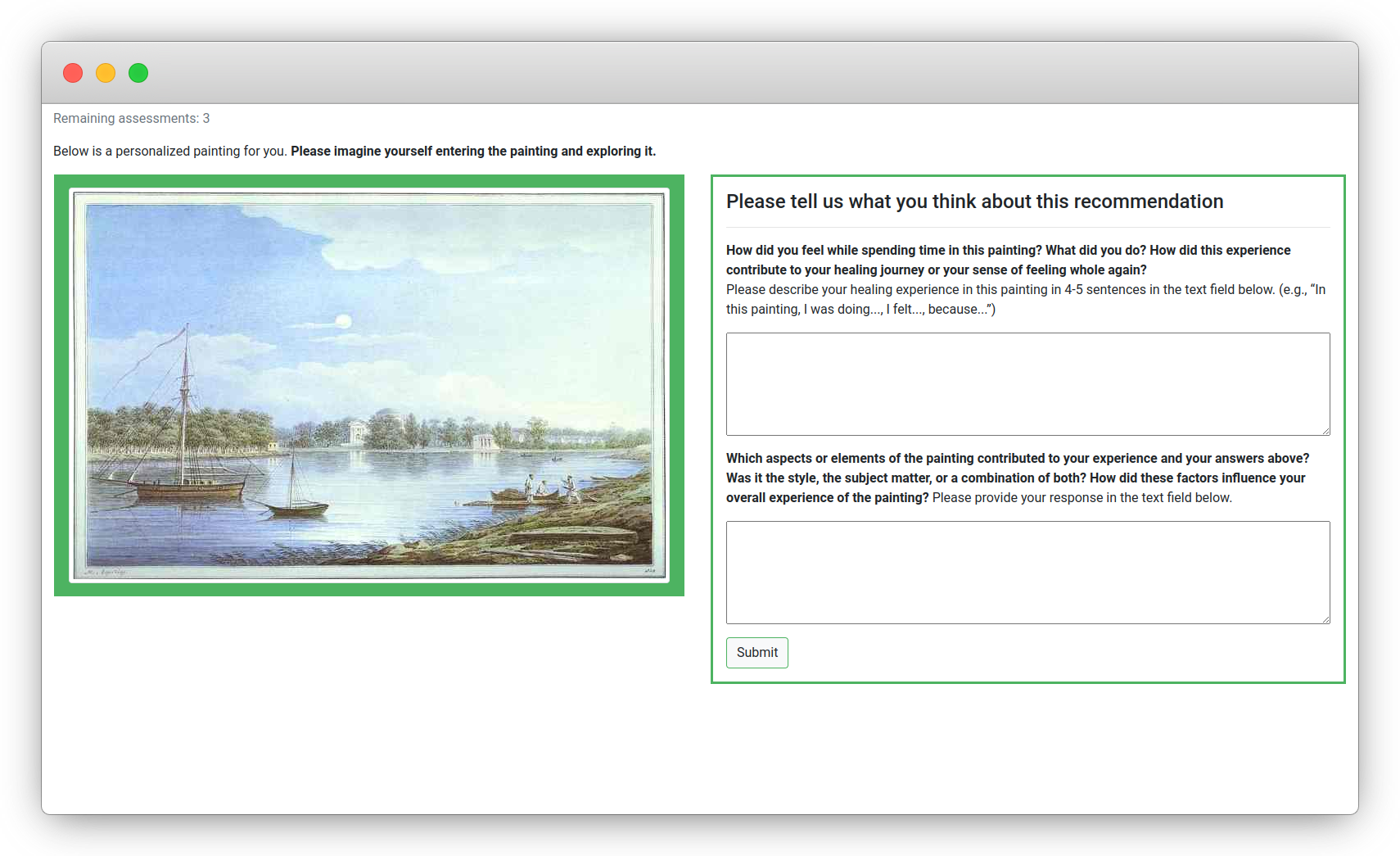}
    }
    \subfloat[Post-study questions\label{fig:web:after}]{
        \includegraphics[width=\w]{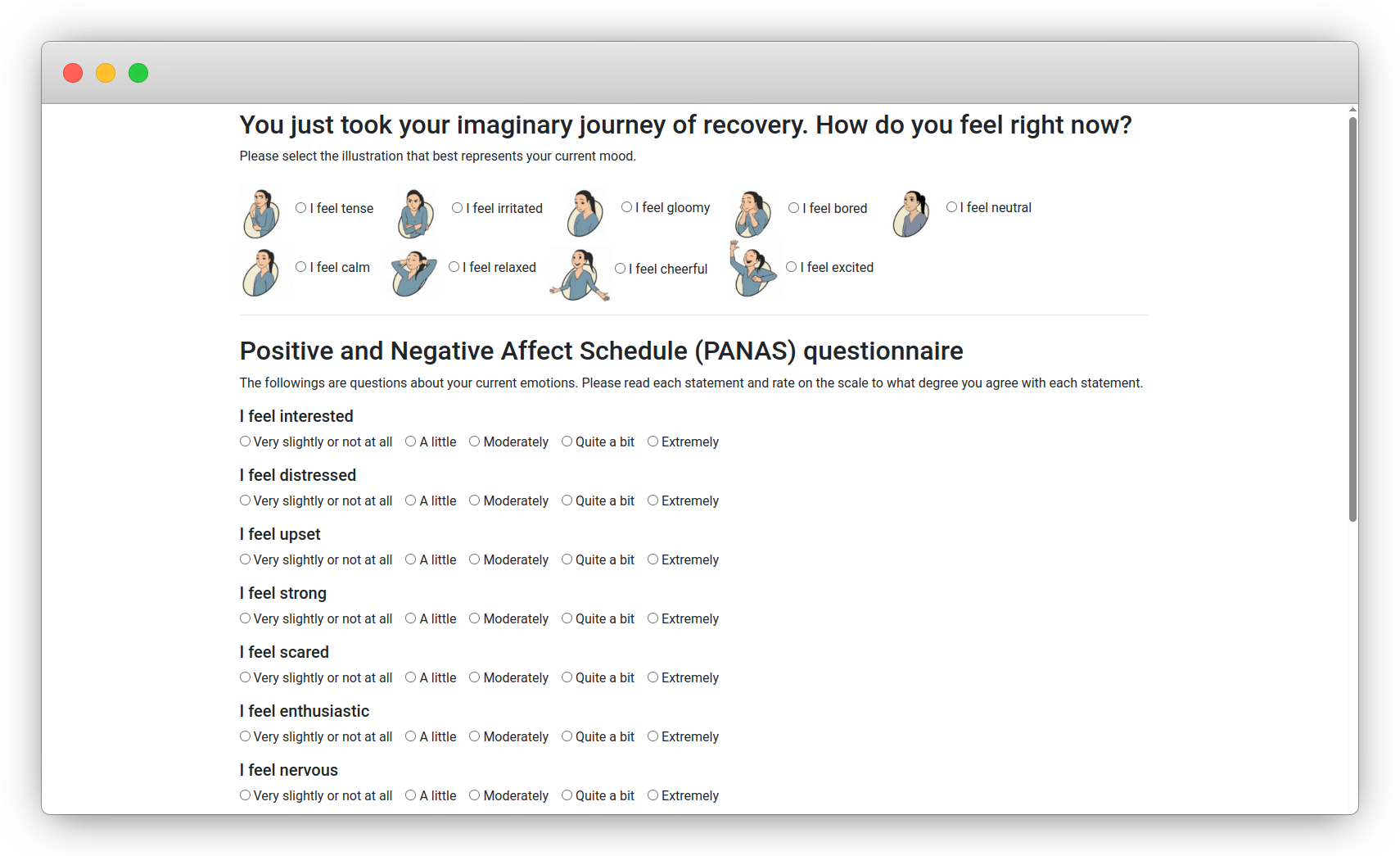}
    }
    \caption{Screenshots (in chronological order) of the web application we developed for the user study.}
    \Description{Screenshots of the web application. From left to right: pre-study questions, preference elicitation, guided AT, and post-study questions.}
    \label{fig:webapp}
\end{figure*}

\section{User Study}
\label{sec:ustd}

We conducted a large-scale online study to understand the user's perception towards the quality of our CDR strategies 
for psychiatric sequelae rehabilitation therapy, 
and ultimately to assess their efficacy in supporting their healing journey. 
Hence, we designed a guided AT study that follows the setup 
proposed in SOTA VA RecSys for digital AT~\cite{yilma2024artful, yilma2025ai}. 
In addition to the three CDR engines, we developed a SOTA Visual engine
that produces recommendations based on elicited painting preferences
using cosine similarity over ResNet50-extracted features.

\subsection{Materials}

We curated the music and painting samples from DEAM and WikiArt
to ensure that they were suitable for conducting guided AT sessions.
Music and paintings were retained if their valence score exceeded 0.1, 
ensuring positive emotionality as per prior therapeutic studies~\cite{roy2008emotional, campos2015emotional},
and arousal was either $\leq -0.1$ or $\geq 0.1$, 
excluding neutral samples to enhance emotional engagement~\cite{sandstrom2010music, wheeler2011musically}. 

The curated music set supported preference elicitation, 
while paintings were matched with a set of nature-based artworks 
that were previously validated to support AT~\cite{yilma2024artful}. 
This allowed us to identify paintings in WikiArt closely matching those used in digital AT. 
Further, to ensure a safe deployment, we conducted a pilot test with a registered art therapist\footnote{\url{https://www.arttherapyfederation.eu/}} 
to manually go through the curated paintings and remove potentially harmful or unsuitable paintings, 
yielding a final $239 \times 63$ similarity matrix (239 music tracks, 63 paintings).

\subsection{Participants}

We recruited 200 people who had psychiatric sequelae
via the Prolific crowdsourcing platform.\footnote{\url{https://www.prolific.com/}}
All participants indicated that they were officially diagnosed 
with any of the following mental health conditions:
Anxiety (165 participants), 
Depression (162), 
Attention Deficit Hyperactivity Disorder (70), 
Bipolar Disorder (45), 
Eating Disorder (36), 
Post-traumatic Stress Disorder (28), 
among other conditions.
We also administered the Patient Health Questionnaire-4 (PHQ-4)~\cite{lowe20104} 
to collect explicit signs of anxiety and depression.
All participants showed symptoms for both of these conditions.
Participants also indicated to have been through any of the following treatments:
Psychological therapy (188),
Medication (137),
Self-care (71),
Hospital and residential treatment programs (47),
and Brain-stimulation treatments (27).
Furthermore, we enforced other criteria for eligibility:  
being fluent in English, minimum approval rate of 100\% in previous crowdsourcing studies in Prolific, 
and being active in the last 90 days.

Our recruited participants (100 female, 100 male) were aged 31.4 years (SD=10.7) 
and could complete the study only once. 
Most of them lived in South Africa (85), Poland (21), United Kingdom (16), and Canada (12).
Participants had been through 
planned or emergency surgical procedures (88),
traumatic experience such as injuries from an accident or a fall (68),
or medical conditions such as pneumonia, heart failure, or stroke (57).
Participants had been in a general ward or medical/surgical unit (85), ICU (73), or in an Emergency Room (54).
For most participants, the duration of their stay in a hospital was 
less than a week (89), between 1 to 2 weeks (47), or between 2 to 4 weeks (45).
The study took a mean time of 35\,min (SD=15) to complete.
Participants were paid an equivalent hourly wage of \$12/h.

\subsection{Apparatus}

We designed a web application (\autoref{fig:webapp}) that first assessed baseline affective states,
then elicited music and painting preferences,\footnote{Paintings were necessary for the Visual engine only.}
then facilitated a guided AT session,
and finally assessed post-test affective states.
We indicated in Prolific that participants had to use a desktop-based computer,
to ensure they would be in a comfortable environment,
and that the study would require them to use headphones.

\subsection{Design}

We deployed the three CDR engines (Mozart, Haydn, and Salieri) 
together with the SOTA baseline VA RecSys engine (Visual). 
Each participant was only exposed to one of the four engines (between-subjects design) 
and then went through a session of guided AT comprising three paintings 
(the top-3 recommendations from each engine, based on the elicited user preferences, see \autoref{algo:recommendation}).
Each group comprised a gender-balanced set of 50 participants.

\begin{figure*}[!ht]
    \centering
    \def\w{0.15\linewidth}
    
    \includegraphics[width=\w]{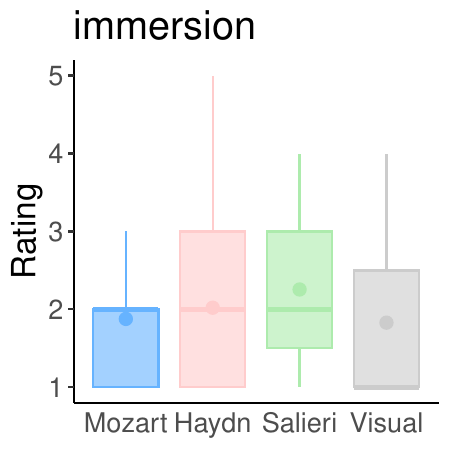} \hfil
    \includegraphics[width=\w]{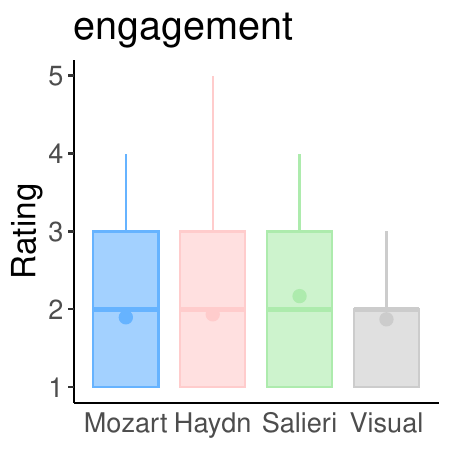} \hfil
    \includegraphics[width=\w]{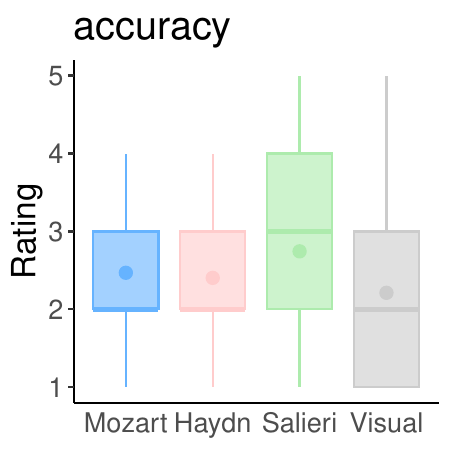} \hfil
    \includegraphics[width=\w]{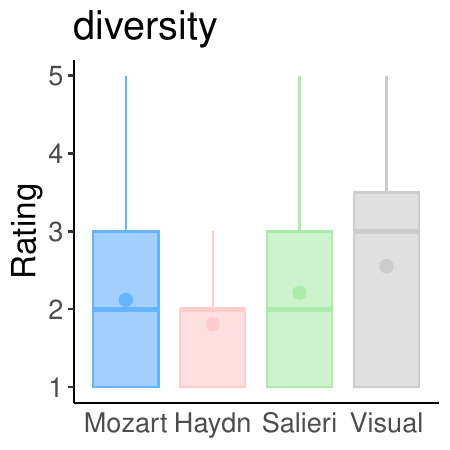} \hfil
    \includegraphics[width=\w]{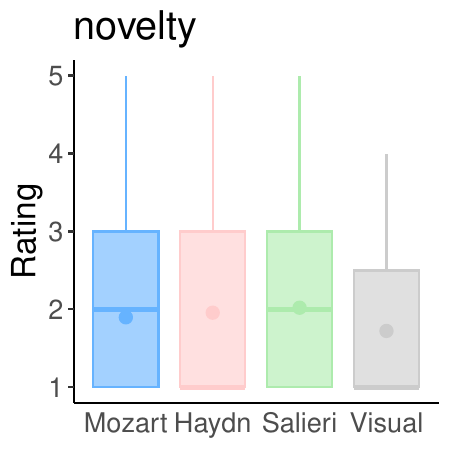} \hfil
    \includegraphics[width=\w]{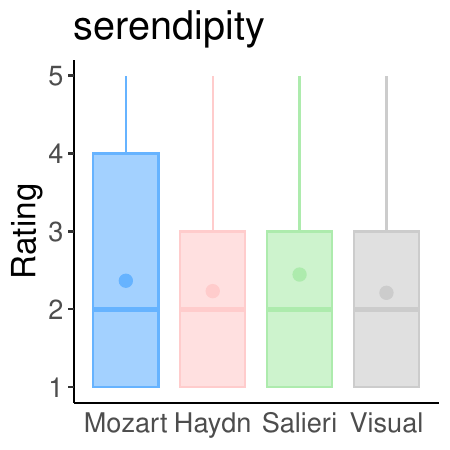} 

    \caption{
        Distribution of ratings of our user-centric recommendation quality metrics.
        Dots denote mean values.
    }
    \Description{
        Statistical plots of user ratings according to each of our user-centric metrics of recommendation quality:
        accuracy, diversity, novelty, serendipity, immersion, engagement. 
        All proposed CDR approaches performed similarly, 
        which in turn outperformed the SOTA approach in most cases.
    }
    \label{fig:prolific-ratings-overall}
\end{figure*}

\begin{figure*}[!ht]
    \centering
    \includegraphics[width=\linewidth]{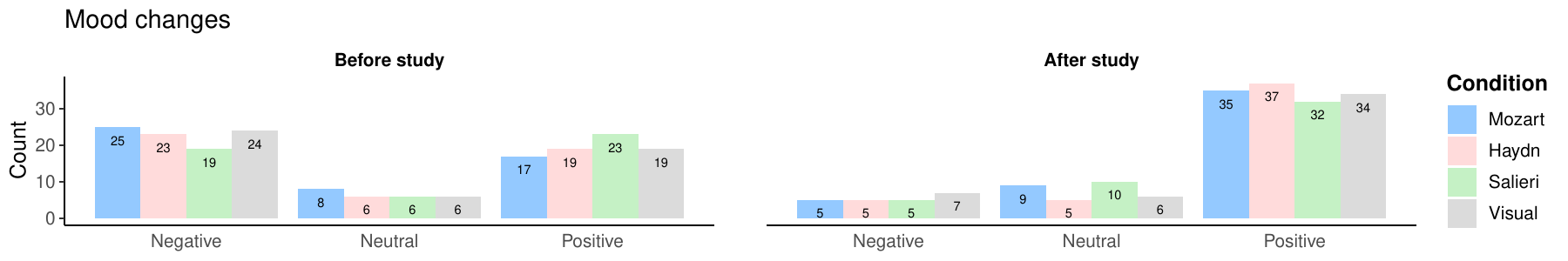}
    \caption{
        Mood improvement comparison before and after art therapy.
    }
    \Description{
        Bar plot showing mood scores according the four groups under study.
        All groups showed a comparable positive improvement after Art therapy.
    }
    \label{fig:mood_improvement}
\end{figure*}

\begin{figure*}[!ht]
    \centering
    \includegraphics[width=\linewidth]{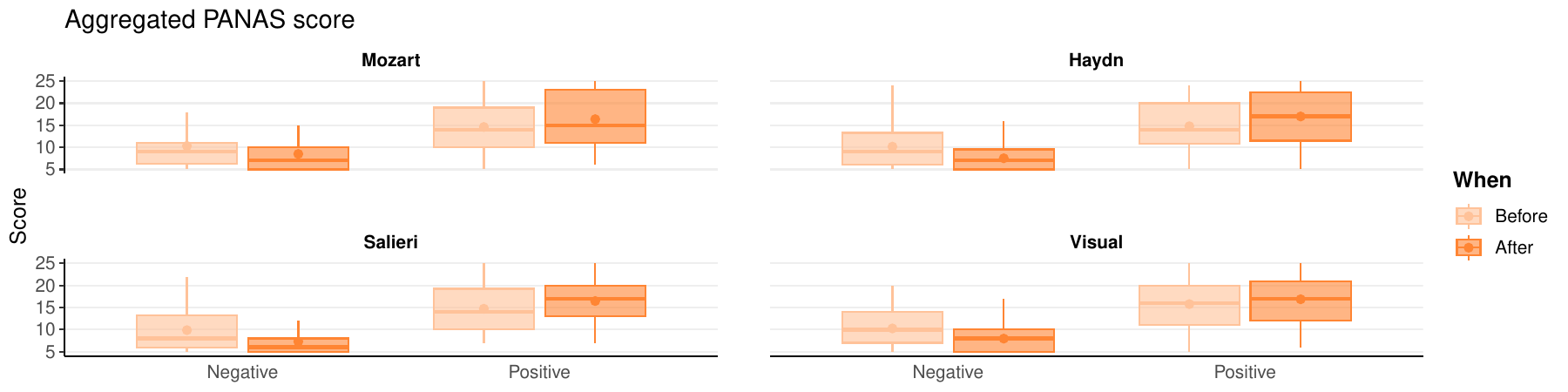}
    \caption{
        Emotion score changes according to the Positive Affect Negative Affect Schedule (PANAS) scale.
    }
    \Description{
        Statistical plots of the aggregated PANAS score according to each dimension. 
        All proposed CDR approaches performed similarly, which in turn outperformed the SOTA approach.
    }
    \label{fig:panas-aggregated}
\end{figure*}

\subsection{Procedure}

We elicited user preferences by providing a set of 11 music tracks and 11 paintings,
randomly sampled from DEAM and WikiArt. 
Participants had to rate each item in a 1--5 scale. 
One music track and one painting were attentional checks, 
which had to be rated as 1 and were not used by any engine. 
Participants had to listen each music track for 15 seconds.

We assessed pre-test and post-test affective states using
the Pick-A-Mood tool~\cite{desmet2016mood} 
and the short version of the Positive and Negative Affect Schedule (PANAS) scale~\cite{watson1988development},
which considers 5 positive, 5 negative, and 1 neutral item 
to assess the affective state of participants before and after guided AT. 

The guided AT session comprised three paintings that were recommended to each participant based on the RecSys engine they were assigned to.
Participants were encouraged to engage with the paintings:
\textit{``Imagine yourself entering the painting and exploring it. How did you feel while spending time in this painting?''}.
They were prompted to reflect on their healing experience and describe it in a few sentences.
They were also prompted to describe which aspects of the painting contributed most to their experience.

After the AT session, in addition to the affective states self-assessment, 
participants were asked to rate the recommendations in a 5-point Likert scale
according to widely accepted proxies of recommendation quality~\cite{pu2011user, yilma2024artful}:
\begin{description}
    \item[Accuracy:] To what degree did the recommended paintings match your selected personal preference?
    \item[Diversity:] To what extent did you find the recommended paintings diverse?
    \item[Novelty:] To what extent did the recommendations help you discover paintings you didn’t know before?
    \item[Serendipity:] To what extent did you find surprisingly interesting paintings among the recommendations?
    \item[Immersion:] How much did the recommended paintings contribute to your sense of immersion, making you feel deeply involved or absorbed in the artwork?
    \item[Engagement:] To what extent did the recommended paintings contribute to your feeling of engagement, capturing your attention and generating a sense of involvement or interest?
\end{description}

\subsection{Results}

Our analysis is structured in terms of 
user-centric metrics of recommendation quality,
emotional changes post-therapy, and user reflections.

\subsubsection{Analysis of recommendation quality measures}

We used a linear mixed-effects (LME) model where participants are considered random effects.\footnote{An LME model is appropriate because ratings are discrete and have a natural order.}
We fitted one LME model per user-centric metric 
and computed the estimated marginal means for specified factors.
We then ran pairwise comparisons (also known as \emph{contrasts})
with Bonferroni-Holm correction to guard against multiple comparisons.

\autoref{fig:prolific-ratings-overall} shows the distributions of user ratings
for the user-centric metrics of recommendation quality.
We did not find statistically significant differences between engines,
suggesting that all engines were rated similarly by our participants.
The only exception is Visual outperforming Haydn in terms of diversity ($p = .007$) 
with a moderate effect size ($r=0.2$).

\subsubsection{Analysis of changes in mood}

\autoref{fig:mood_improvement} shows the mood changes before and after therapy,
for the four recommendation groups we have considered in our study. 
In all groups, a mood enhancement effect of guided art therapy was observed. 
When comparing it to the baseline, where many participants were in a negative mood (46.6\%), 
after guided art therapy the majority of participants reported being in a positive mood (72.6\%), 
with only a minority remaining in negative (11.5\%) or neutral (15.7\%) mood.
The differences between approaches were not statistically significant:
$\chi^2{(3, N=195)} = 1.08, p = .780, \phi = 0.075$.

\autoref{fig:panas-aggregated} shows the change in scores after therapy,
aggregated according to the ten items of the PANAS scale.
Differences between groups were not statistically significant in any case, with small to moderate effect sizes.

\subsubsection{Analysis of user reflections}

\begin{figure}[!ht]
    \centering
    \includegraphics[width=0.8\columnwidth]{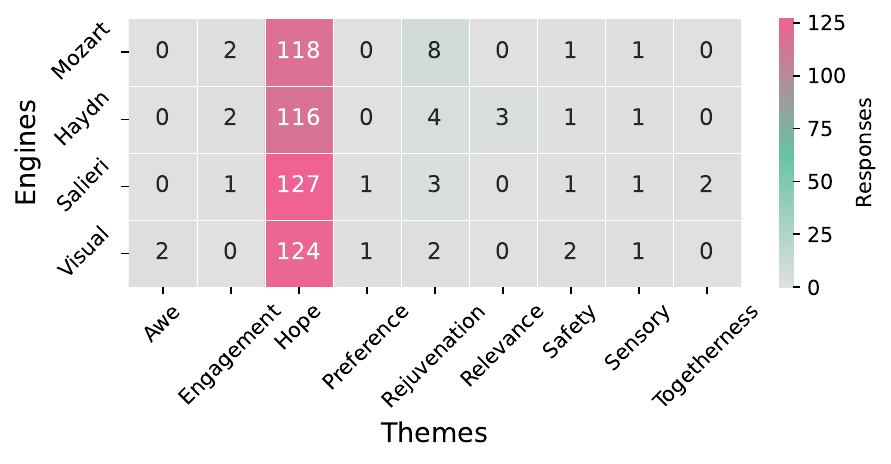}
    \caption{Heatmap of healing concepts identified by recommendation engines using guided topic modeling.}
    \Description{A heatmap comparing the number of participant responses assigned to 10 healing themes across four recommendation engines: Visual, Haydn, Mozart, and Salieri. The y-axis lists the healing themes: Awe, Engagement, Hope, Preference, Rejuvenation, Relevance, Safety, Sensory, Togetherness, and Escape. The x-axis lists the recommendation engines: Visual, Haydn, Mozart, and Salieri. Cell colors range from light yellow (low counts) to dark blue (high counts), representing the number of responses, with exact counts annotated in each cell. Hope dominates across all engines, with Salieri at 122, Visual at 112, Mozart at 106, and Haydn at 104 responses. Rejuvenation follows, with Mozart at 17, Visual at 14, Haydn at 11, and Salieri at 9. Other themes like Engagement (Mozart: 6, Visual: 8) and Relevance (Mozart: 14, Visual: 5) show limited presence, while Awe, Safety, Sensory, and Togetherness have 1–3 responses each. A color bar on the right indicates the scale of responses, labeled "Number of Responses." The background is white with thin lines between cells.}
    \label{fig:healing_concepts}
\end{figure}

We conducted a thematic analysis to identify the most salient healing themes
from the user reflections, which were submitted as free-form comments during the AT session.
Starting from a list of 10 themes (with sample quotes from participants)
that follow established therapeutic practice~\cite{yilma2025ai},
we seeded BERTopic~\cite{grootendorst2022bertopic} 
with keywords derived from each theme’s description and sample quote.
For example, for the Safety theme, the following keywords were identified: 
"safe", "secure", "protected", "calm", "familiar".
Then, each user reflection towards a particular painting
was assigned a dominant theme based on topic coherence, 
with mappings refined by examining the top words per topic. 
The resulting theme frequencies are reported in \autoref{fig:healing_concepts}.

Notably, all engines predominantly elicited Hope, 
with Salieri showing the highest focus (127 responses), 
followed by Visual (124), Mozart (118), and Haydn (116). 
Again, compared to the SOTA image-based engine, 
the music-based approaches demonstrate a competitive therapeutic profile, 
even outperforming SOTA in most user-centric metrics of recommendation quality,
suggesting that music-based preference elicitation methods 
are an effective alternative in personalized AT.

\section{Discussion}

We have made the first step in digital AT personalization 
by introducing affect-aware music-driven preference elicitation. 
Taken together, our CDR algorithms (Mozart, Haydn, and Salieri)\footnote{Mozart, Haydn, and Salieri represent not just classical music composers but contrasting philosophies (brilliance, balance, and rivalry). These names were chosen to mirror the unique design principles  behind each recommendation method.} 
unlock new possibilities, mapping music’s affective cues to therapeutic VA recommendations. 
The findings from our study reveal distinct strengths of our approaches.  
For example, Salieri excels in promoting Hope (127 responses, \autoref{fig:healing_concepts}) 
while Haydn delivers interpretable affective alignment. 
This suggests that music-driven elicitation offers focused emotional resonance. 
Furthermore, mood improvements (46.6\% negative to 72.6\% positive post-therapy, \autoref{fig:mood_improvement}) 
across all proposed CDR algorithms also signify music’s broader emotional appeal.  
All in all, these findings suggest CDR holds potential 
to open new avenues for digital AT via multimodal frameworks. 
For example, AT systems could connect to music streaming applications like Spotify or Dizzer, 
thus leveraging the users' listening history or playlists to derive therapeutic recommendations.

\section{Limitations and Future Work}

A primary limitation in this work is the scarcity of 
therapeutically curated painting collections with reliable affective labels. 
Prior work~\cite{yilma2025ai, yilma2024artful} 
highlighted the potential harm of unfiltered visual stimuli in digital AT, 
necessitating therapist intervention to safeguard recommendations. 
We align with this view, yet the lack of parallel datasets, 
affectively labeled and within therapeutic valence-arousal ranges, 
poses a significant challenge to scaling music-driven VA recommendations.
Future efforts could address this through crowdsourced studies, 
similar to WikiArt Emotions, to curate robust therapeutic datasets, 
fostering a research space where affect-aware CDR algorithms can excel. 
Furthermore, the lack of significant differences between systems may be attributed to stimulus homogeneity, 
as a result of filtered samples implemented as safety measures to prevent the exposure of patients to negative stimuli. 
We believe longitudinal studies to further validate the therapeutic impacts of these approaches are exciting avenues for future work.

 \section{Conclusion}
We have introduced affect-aware cross-domain recommendation for digital art therapy by leveraging music preference elicitation to personalize visual art recommendations. Our family of methods (Mozart, Haydn, Salieri) bridge the music-art gap, enhancing emotional engagement through cross-modal affective transfer. Our study showed these approaches match state-of-the-art visual baselines in metrics like accuracy and diversity while significantly improving mood and positive affect. This work advances RecSys by addressing single-domain limitations, and we hope to inspire multimodal RecSys for well-being.

\begin{acks}
Research supported by the Pathfinder program of the European Innovation Council (grant 101071147)
and the Horizon 2020 FET program of the European Union (grant CHIST-ERA-20-BCI-001).
\end{acks}

\bibliographystyle{ACM-Reference-Format}
\bibliography{sample-base}


\begin{thebibliography}{54}


\ifx \showCODEN    \undefined \def \showCODEN     #1{\unskip}     \fi
\ifx \showDOI      \undefined \def \showDOI       #1{#1}\fi
\ifx \showISBNx    \undefined \def \showISBNx     #1{\unskip}     \fi
\ifx \showISBNxiii \undefined \def \showISBNxiii  #1{\unskip}     \fi
\ifx \showISSN     \undefined \def \showISSN      #1{\unskip}     \fi
\ifx \showLCCN     \undefined \def \showLCCN      #1{\unskip}     \fi
\ifx \shownote     \undefined \def \shownote      #1{#1}          \fi
\ifx \showarticletitle \undefined \def \showarticletitle #1{#1}   \fi
\ifx \showURL      \undefined \def \showURL       {\relax}        \fi
\providecommand\bibfield[2]{#2}
\providecommand\bibinfo[2]{#2}
\providecommand\natexlab[1]{#1}
\providecommand\showeprint[2][]{arXiv:#2}

\bibitem[Abbing et~al\mbox{.}(2018)]%
        {abbing2018effectiveness}
\bibfield{author}{\bibinfo{person}{Annemarie Abbing}, \bibinfo{person}{Anne Ponstein}, \bibinfo{person}{Susan van Hooren}, \bibinfo{person}{Leo de Sonneville}, \bibinfo{person}{Hanna Swaab}, {and} \bibinfo{person}{Erik Baars}.} \bibinfo{year}{2018}\natexlab{}.
\newblock \showarticletitle{The effectiveness of art therapy for anxiety in adults: A systematic review of randomised and non-randomised controlled trials}.
\newblock \bibinfo{journal}{\emph{PloS one}} \bibinfo{volume}{13}, \bibinfo{number}{12} (\bibinfo{year}{2018}), \bibinfo{pages}{e0208716}.
\newblock


\bibitem[Aslan et~al\mbox{.}(2022)]%
        {aslan2022recognizing}
\bibfield{author}{\bibinfo{person}{Sinem Aslan}, \bibinfo{person}{Giovanna Castellano}, \bibinfo{person}{Vincenzo Digeno}, \bibinfo{person}{Giuseppe Migailo}, \bibinfo{person}{Raffaele Scaringi}, {and} \bibinfo{person}{Gennaro Vessio}.} \bibinfo{year}{2022}\natexlab{}.
\newblock \showarticletitle{Recognizing the emotions evoked by artworks through visual features and knowledge graph-embeddings}. In \bibinfo{booktitle}{\emph{International Conference on Image Analysis and Processing}}. Springer, \bibinfo{pages}{129--140}.
\newblock


\bibitem[Barnett and Vasiu(2024)]%
        {barnett2024arts}
\bibfield{author}{\bibinfo{person}{Kelly~Sarah Barnett} {and} \bibinfo{person}{Fabian Vasiu}.} \bibinfo{year}{2024}\natexlab{}.
\newblock \showarticletitle{How the arts heal: a review of the neural mechanisms behind the therapeutic effects of creative arts on mental and physical health}.
\newblock \bibinfo{journal}{\emph{Frontiers in behavioral neuroscience}}  \bibinfo{volume}{18} (\bibinfo{year}{2024}), \bibinfo{pages}{1422361}.
\newblock


\bibitem[Blei et~al\mbox{.}(2003)]%
        {blei2003latent}
\bibfield{author}{\bibinfo{person}{David~M Blei}, \bibinfo{person}{Andrew~Y Ng}, {and} \bibinfo{person}{Michael~I Jordan}.} \bibinfo{year}{2003}\natexlab{}.
\newblock \showarticletitle{Latent dirichlet allocation}.
\newblock \bibinfo{journal}{\emph{Journal of machine Learning research}} \bibinfo{volume}{3}, \bibinfo{number}{Jan} (\bibinfo{year}{2003}), \bibinfo{pages}{993--1022}.
\newblock


\bibitem[Bokoch et~al\mbox{.}(2025)]%
        {bokoch2025scoping}
\bibfield{author}{\bibinfo{person}{Rebecca Bokoch}, \bibinfo{person}{Noah Hass-Cohen}, \bibinfo{person}{April Espinoza}, \bibinfo{person}{Tyler O’Reilly}, {and} \bibinfo{person}{Elad Levi}.} \bibinfo{year}{2025}\natexlab{}.
\newblock \showarticletitle{A scoping review of integrated arts therapies and neuroscience research}.
\newblock \bibinfo{journal}{\emph{Frontiers in Psychology}}  \bibinfo{volume}{16} (\bibinfo{year}{2025}), \bibinfo{pages}{1569609}.
\newblock


\bibitem[Brancatisano et~al\mbox{.}(2020)]%
        {brancatisano2020music}
\bibfield{author}{\bibinfo{person}{Olivia Brancatisano}, \bibinfo{person}{Amee Baird}, {and} \bibinfo{person}{William~Forde Thompson}.} \bibinfo{year}{2020}\natexlab{}.
\newblock \showarticletitle{Why is music therapeutic for neurological disorders? The Therapeutic Music Capacities Model}.
\newblock \bibinfo{journal}{\emph{Neuroscience \& Biobehavioral Reviews}}  \bibinfo{volume}{112} (\bibinfo{year}{2020}), \bibinfo{pages}{600--615}.
\newblock


\bibitem[Campos-Bueno et~al\mbox{.}(2015)]%
        {campos2015emotional}
\bibfield{author}{\bibinfo{person}{Jos{\'e}~J Campos-Bueno}, \bibinfo{person}{Octavio DeJuan-Ayala}, \bibinfo{person}{Pedro Montoya}, {and} \bibinfo{person}{Niels Birbaumer}.} \bibinfo{year}{2015}\natexlab{}.
\newblock \showarticletitle{Emotional dimensions of music and painting and their interaction}.
\newblock \bibinfo{journal}{\emph{The Spanish journal of psychology}}  \bibinfo{volume}{18} (\bibinfo{year}{2015}), \bibinfo{pages}{E54}.
\newblock


\bibitem[Choi et~al\mbox{.}(2024)]%
        {choi2024accelerated}
\bibfield{author}{\bibinfo{person}{Suvin Choi}, \bibinfo{person}{Jong-Ik Park}, \bibinfo{person}{Cheol-Ho Hong}, \bibinfo{person}{Sang-Gue Park}, {and} \bibinfo{person}{Sang-Cheol Park}.} \bibinfo{year}{2024}\natexlab{}.
\newblock \showarticletitle{Accelerated construction of stress relief music datasets using CNN and the Mel-scaled spectrogram}.
\newblock \bibinfo{journal}{\emph{PloS one}} \bibinfo{volume}{19}, \bibinfo{number}{5} (\bibinfo{year}{2024}), \bibinfo{pages}{e0300607}.
\newblock


\bibitem[Chowdhury et~al\mbox{.}(2019)]%
        {chowdhury-2019-towards}
\bibfield{author}{\bibinfo{person}{Shreyan Chowdhury}, \bibinfo{person}{Andreu Vall}, \bibinfo{person}{Verena Haunschmid}, {and} \bibinfo{person}{Gerhard Widmer}.} \bibinfo{year}{2019}\natexlab{}.
\newblock \showarticletitle{Towards Explainable Music Emotion Recognition: The Route via Mid-level Features}. In \bibinfo{booktitle}{\emph{Proceedings of the 20th International Society for Music Information Retrieval Conference, {ISMIR} 2019, Delft, The Netherlands, November 4-8, 2019}}, \bibfield{editor}{\bibinfo{person}{Arthur Flexer}, \bibinfo{person}{Geoffroy Peeters}, \bibinfo{person}{Juli{\'{a}}n Urbano}, {and} \bibinfo{person}{Anja Volk}} (Eds.). \bibinfo{pages}{237--243}.
\newblock
\urldef\tempurl%
\url{http://archives.ismir.net/ismir2019/paper/000027.pdf}
\showURL{%
\tempurl}


\bibitem[Desmet et~al\mbox{.}(2016)]%
        {desmet2016mood}
\bibfield{author}{\bibinfo{person}{Pieter~MA Desmet}, \bibinfo{person}{Martijn~H Vastenburg}, {and} \bibinfo{person}{Natalia Romero}.} \bibinfo{year}{2016}\natexlab{}.
\newblock \showarticletitle{Mood measurement with Pick-A-Mood: review of current methods and design of a pictorial self-report scale}.
\newblock \bibinfo{journal}{\emph{Journal of Design Research}} \bibinfo{volume}{14}, \bibinfo{number}{3} (\bibinfo{year}{2016}), \bibinfo{pages}{241--279}.
\newblock


\bibitem[Devlin et~al\mbox{.}(2018)]%
        {devlin2018bert}
\bibfield{author}{\bibinfo{person}{Jacob Devlin}, \bibinfo{person}{Ming-Wei Chang}, \bibinfo{person}{Kenton Lee}, {and} \bibinfo{person}{Kristina Toutanova}.} \bibinfo{year}{2018}\natexlab{}.
\newblock \showarticletitle{Bert: Pre-training of deep bidirectional transformers for language understanding}.
\newblock \bibinfo{journal}{\emph{arXiv preprint arXiv:1810.04805}} (\bibinfo{year}{2018}).
\newblock


\bibitem[Di~Vita et~al\mbox{.}(2022)]%
        {di2022psychotherapy}
\bibfield{author}{\bibinfo{person}{Antonella Di~Vita}, \bibinfo{person}{Mario~Augusto Procacci}, \bibinfo{person}{Martina Bellagamba}, \bibinfo{person}{Maria Jacomini}, \bibinfo{person}{Roberta Massicci}, {and} \bibinfo{person}{Maria~Paola Ciurli}.} \bibinfo{year}{2022}\natexlab{}.
\newblock \showarticletitle{Psychotherapy and Art Therapy: A pilot study of group treatment for patients with traumatic brain injury}.
\newblock \bibinfo{journal}{\emph{Journal of health psychology}} \bibinfo{volume}{27}, \bibinfo{number}{4} (\bibinfo{year}{2022}), \bibinfo{pages}{836--846}.
\newblock


\bibitem[Elkahky et~al\mbox{.}(2015)]%
        {elkahky2015multi}
\bibfield{author}{\bibinfo{person}{Ali~Mamdouh Elkahky}, \bibinfo{person}{Yang Song}, {and} \bibinfo{person}{Xiaodong He}.} \bibinfo{year}{2015}\natexlab{}.
\newblock \showarticletitle{A multi-view deep learning approach for cross domain user modeling in recommendation systems}. In \bibinfo{booktitle}{\emph{Proceedings of the 24th international conference on world wide web}}. \bibinfo{pages}{278--288}.
\newblock


\bibitem[Feng and Wang(2025)]%
        {feng2025effect}
\bibfield{author}{\bibinfo{person}{Yingjie Feng} {and} \bibinfo{person}{Mingda Wang}.} \bibinfo{year}{2025}\natexlab{}.
\newblock \showarticletitle{Effect of music therapy on emotional resilience, well-being, and employability: a quantitative investigation of mediation and moderation}.
\newblock \bibinfo{journal}{\emph{BMC psychology}} \bibinfo{volume}{13}, \bibinfo{number}{1} (\bibinfo{year}{2025}), \bibinfo{pages}{47}.
\newblock


\bibitem[Fern{\'a}ndez-Tob{\'\i}as et~al\mbox{.}(2016)]%
        {fernandez2016alleviating}
\bibfield{author}{\bibinfo{person}{Ignacio Fern{\'a}ndez-Tob{\'\i}as}, \bibinfo{person}{Matthias Braunhofer}, \bibinfo{person}{Mehdi Elahi}, \bibinfo{person}{Francesco Ricci}, {and} \bibinfo{person}{Iv{\'a}n Cantador}.} \bibinfo{year}{2016}\natexlab{}.
\newblock \showarticletitle{Alleviating the new user problem in collaborative filtering by exploiting personality information}.
\newblock \bibinfo{journal}{\emph{User Modeling and User-Adapted Interaction}}  \bibinfo{volume}{26} (\bibinfo{year}{2016}), \bibinfo{pages}{221--255}.
\newblock


\bibitem[Grootendorst(2022)]%
        {grootendorst2022bertopic}
\bibfield{author}{\bibinfo{person}{Maarten Grootendorst}.} \bibinfo{year}{2022}\natexlab{}.
\newblock \showarticletitle{BERTopic: Neural topic modeling with a class-based TF-IDF procedure}.
\newblock \bibinfo{journal}{\emph{arXiv preprint arXiv:2203.05794}} (\bibinfo{year}{2022}).
\newblock


\bibitem[Haeyen and Staal(2021)]%
        {haeyen2021imagery}
\bibfield{author}{\bibinfo{person}{Suzanne Haeyen} {and} \bibinfo{person}{Merel Staal}.} \bibinfo{year}{2021}\natexlab{}.
\newblock \showarticletitle{Imagery rehearsal based art therapy: Treatment of post-traumatic nightmares in art therapy}.
\newblock \bibinfo{journal}{\emph{Frontiers in psychology}}  \bibinfo{volume}{11} (\bibinfo{year}{2021}), \bibinfo{pages}{628717}.
\newblock


\bibitem[Hathorn and Nanda(2008)]%
        {hathorn2008guide}
\bibfield{author}{\bibinfo{person}{Kathy Hathorn} {and} \bibinfo{person}{Upali Nanda}.} \bibinfo{year}{2008}\natexlab{}.
\newblock \showarticletitle{A guide to evidence-based art}.
\newblock \bibinfo{journal}{\emph{The Center for Health Design}}  \bibinfo{volume}{1} (\bibinfo{year}{2008}), \bibinfo{pages}{1--23}.
\newblock


\bibitem[He et~al\mbox{.}(2016)]%
        {he2016deep}
\bibfield{author}{\bibinfo{person}{Kaiming He}, \bibinfo{person}{Xiangyu Zhang}, \bibinfo{person}{Shaoqing Ren}, {and} \bibinfo{person}{Jian Sun}.} \bibinfo{year}{2016}\natexlab{}.
\newblock \showarticletitle{Deep residual learning for image recognition}. In \bibinfo{booktitle}{\emph{Proceedings of the IEEE conference on computer vision and pattern recognition}}. \bibinfo{pages}{770--778}.
\newblock


\bibitem[Hirakawa et~al\mbox{.}(2021)]%
        {hirakawa2021cross}
\bibfield{author}{\bibinfo{person}{Taisei Hirakawa}, \bibinfo{person}{Keisuke Maeda}, \bibinfo{person}{Takahiro Ogawa}, \bibinfo{person}{Satoshi Asamizu}, {and} \bibinfo{person}{Miki Haseyama}.} \bibinfo{year}{2021}\natexlab{}.
\newblock \showarticletitle{Cross-domain recommendation method based on multi-layer graph analysis with visual information}. In \bibinfo{booktitle}{\emph{2021 IEEE International Conference on Image Processing (ICIP)}}. IEEE, \bibinfo{pages}{2688--2692}.
\newblock


\bibitem[Hu et~al\mbox{.}(2021)]%
        {hu2021art}
\bibfield{author}{\bibinfo{person}{Jingxuan Hu}, \bibinfo{person}{Jinhuan Zhang}, \bibinfo{person}{Liyu Hu}, \bibinfo{person}{Haibo Yu}, {and} \bibinfo{person}{Jinping Xu}.} \bibinfo{year}{2021}\natexlab{}.
\newblock \showarticletitle{Art therapy: a complementary treatment for mental disorders}.
\newblock \bibinfo{journal}{\emph{Frontiers in psychology}}  \bibinfo{volume}{12} (\bibinfo{year}{2021}), \bibinfo{pages}{686005}.
\newblock


\bibitem[Iordanis(2021)]%
        {iordanis2021emotion}
\bibfield{author}{\bibinfo{person}{Papadopoulos~Stefanos Iordanis}.} \bibinfo{year}{2021}\natexlab{}.
\newblock \showarticletitle{Emotion-aware music recommendation systems}.
\newblock  (\bibinfo{year}{2021}).
\newblock


\bibitem[Jiao(2025)]%
        {jiao2025advancing}
\bibfield{author}{\bibinfo{person}{Dian Jiao}.} \bibinfo{year}{2025}\natexlab{}.
\newblock \showarticletitle{Advancing personalized digital therapeutics: integrating music therapy, brainwave entrainment methods, and AI-driven biofeedback}.
\newblock \bibinfo{journal}{\emph{Frontiers in Digital Health}}  \bibinfo{volume}{7} (\bibinfo{year}{2025}), \bibinfo{pages}{1552396}.
\newblock


\bibitem[Juslin and V{\"a}stfj{\"a}ll(2008)]%
        {juslin2008emotional}
\bibfield{author}{\bibinfo{person}{Patrik~N. Juslin} {and} \bibinfo{person}{Daniel V{\"a}stfj{\"a}ll}.} \bibinfo{year}{2008}\natexlab{}.
\newblock \showarticletitle{Emotional responses to music: The need to consider underlying mechanisms}.
\newblock \bibinfo{journal}{\emph{Behavioral and Brain Sciences}} \bibinfo{volume}{31}, \bibinfo{number}{5} (\bibinfo{year}{2008}), \bibinfo{pages}{559--575}.
\newblock
\urldef\tempurl%
\url{https://doi.org/10.1017/S0140525X08005293}
\showDOI{\tempurl}


\bibitem[Koelsch(2014)]%
        {koelsch2014brain}
\bibfield{author}{\bibinfo{person}{Stefan Koelsch}.} \bibinfo{year}{2014}\natexlab{}.
\newblock \showarticletitle{Brain correlates of music-evoked emotions}.
\newblock \bibinfo{journal}{\emph{Nature reviews neuroscience}} \bibinfo{volume}{15}, \bibinfo{number}{3} (\bibinfo{year}{2014}), \bibinfo{pages}{170--180}.
\newblock


\bibitem[Ladas et~al\mbox{.}(2024)]%
        {ladas2024harmony}
\bibfield{author}{\bibinfo{person}{AI Ladas}, \bibinfo{person}{T Gravalas}, \bibinfo{person}{C Katsoridou}, {and} \bibinfo{person}{CA Frantzidis}.} \bibinfo{year}{2024}\natexlab{}.
\newblock \showarticletitle{Harmony in the brain: A narrative review on the shared neural substrates of emotion regulation and creativity}.
\newblock \bibinfo{journal}{\emph{Brain Organoid and Systems Neuroscience Journal}}  \bibinfo{volume}{2} (\bibinfo{year}{2024}), \bibinfo{pages}{81--91}.
\newblock


\bibitem[Lee et~al\mbox{.}(2020)]%
        {lee2020crossing}
\bibfield{author}{\bibinfo{person}{Cheng-Che Lee}, \bibinfo{person}{Wan-Yi Lin}, \bibinfo{person}{Yen-Ting Shih}, \bibinfo{person}{Pei-Yi Kuo}, {and} \bibinfo{person}{Li Su}.} \bibinfo{year}{2020}\natexlab{}.
\newblock \showarticletitle{Crossing you in style: Cross-modal style transfer from music to visual arts}. In \bibinfo{booktitle}{\emph{Proceedings of the 28th ACM international conference on multimedia}}. \bibinfo{pages}{3219--3227}.
\newblock


\bibitem[Li et~al\mbox{.}(2022)]%
        {li2022blip}
\bibfield{author}{\bibinfo{person}{Junnan Li}, \bibinfo{person}{Dongxu Li}, \bibinfo{person}{Caiming Xiong}, {and} \bibinfo{person}{Steven Hoi}.} \bibinfo{year}{2022}\natexlab{}.
\newblock \showarticletitle{Blip: Bootstrapping language-image pre-training for unified vision-language understanding and generation}.
\newblock \bibinfo{journal}{\emph{arXiv preprint arXiv:2201.12086}} (\bibinfo{year}{2022}).
\newblock


\bibitem[Li et~al\mbox{.}(2023)]%
        {li2023mert}
\bibfield{author}{\bibinfo{person}{Yizhi Li}, \bibinfo{person}{Ruibin Yuan}, \bibinfo{person}{Ge Zhang}, \bibinfo{person}{Yinghao Ma}, \bibinfo{person}{Xingran Chen}, \bibinfo{person}{Hanzhi Yin}, \bibinfo{person}{Chenghao Xiao}, \bibinfo{person}{Chenghua Lin}, \bibinfo{person}{Anton Ragni}, \bibinfo{person}{Emmanouil Benetos}, {et~al\mbox{.}}} \bibinfo{year}{2023}\natexlab{}.
\newblock \showarticletitle{Mert: Acoustic music understanding model with large-scale self-supervised training}.
\newblock \bibinfo{journal}{\emph{arXiv preprint arXiv:2306.00107}} (\bibinfo{year}{2023}).
\newblock


\bibitem[Liu and Liu(2022)]%
        {liu2022human}
\bibfield{author}{\bibinfo{person}{Bojun Liu} {and} \bibinfo{person}{Bohong Liu}.} \bibinfo{year}{2022}\natexlab{}.
\newblock \showarticletitle{Human-Centric C ross-Domain T ransfer N etwork for M usic Recommendation}. In \bibinfo{booktitle}{\emph{International Advanced Computing Conference}}. Springer, \bibinfo{pages}{407--414}.
\newblock


\bibitem[L{\"o}we et~al\mbox{.}(2010)]%
        {lowe20104}
\bibfield{author}{\bibinfo{person}{Bernd L{\"o}we}, \bibinfo{person}{Inka Wahl}, \bibinfo{person}{Matthias Rose}, \bibinfo{person}{Carsten Spitzer}, \bibinfo{person}{Heide Glaesmer}, \bibinfo{person}{Katja Wingenfeld}, \bibinfo{person}{Antonius Schneider}, {and} \bibinfo{person}{Elmar Br{\"a}hler}.} \bibinfo{year}{2010}\natexlab{}.
\newblock \showarticletitle{A 4-item measure of depression and anxiety: validation and standardization of the Patient Health Questionnaire-4 (PHQ-4) in the general population}.
\newblock \bibinfo{journal}{\emph{Journal of affective disorders}} \bibinfo{volume}{122}, \bibinfo{number}{1-2} (\bibinfo{year}{2010}), \bibinfo{pages}{86--95}.
\newblock


\bibitem[Malchiodi(2011)]%
        {malchiodi2011handbook}
\bibfield{author}{\bibinfo{person}{Cathy~A Malchiodi}.} \bibinfo{year}{2011}\natexlab{}.
\newblock \bibinfo{booktitle}{\emph{Handbook of art therapy}}.
\newblock \bibinfo{publisher}{Guilford Press}.
\newblock


\bibitem[Man et~al\mbox{.}(2017)]%
        {man2017cross}
\bibfield{author}{\bibinfo{person}{Tong Man}, \bibinfo{person}{Huawei Shen}, \bibinfo{person}{Xiaolong Jin}, {and} \bibinfo{person}{Xueqi Cheng}.} \bibinfo{year}{2017}\natexlab{}.
\newblock \showarticletitle{Cross-domain recommendation: An embedding and mapping approach}. In \bibinfo{booktitle}{\emph{IJCAI}}, Vol.~\bibinfo{volume}{17}. \bibinfo{pages}{2464--2470}.
\newblock


\bibitem[Mohammad(2025)]%
        {mohammad2025nrc}
\bibfield{author}{\bibinfo{person}{Saif~M Mohammad}.} \bibinfo{year}{2025}\natexlab{}.
\newblock \showarticletitle{NRC VAD Lexicon v2: Norms for valence, arousal, and dominance for over 55k English terms}.
\newblock \bibinfo{journal}{\emph{arXiv preprint arXiv:2503.23547}} (\bibinfo{year}{2025}).
\newblock


\bibitem[Panwar et~al\mbox{.}(2019)]%
        {panwar2019you}
\bibfield{author}{\bibinfo{person}{Sharaj Panwar}, \bibinfo{person}{Paul Rad}, \bibinfo{person}{Kim-Kwang~Raymond Choo}, {and} \bibinfo{person}{Mehdi Roopaei}.} \bibinfo{year}{2019}\natexlab{}.
\newblock \showarticletitle{Are you emotional or depressed? Learning about your emotional state from your music using machine learning}.
\newblock \bibinfo{journal}{\emph{The Journal of Supercomputing}}  \bibinfo{volume}{75} (\bibinfo{year}{2019}), \bibinfo{pages}{2986--3009}.
\newblock


\bibitem[Patel and Mehta(2024)]%
        {patel2024emotion}
\bibfield{author}{\bibinfo{person}{Varesh Patel} {and} \bibinfo{person}{Khyati Mehta}.} \bibinfo{year}{2024}\natexlab{}.
\newblock \showarticletitle{Emotion-Aware Music Recommendations: Evaluating Custom CNN vs. VGG16 and InceptionV3}. In \bibinfo{booktitle}{\emph{2024 IEEE International Conference on Future Machine Learning and Data Science (FMLDS)}}. IEEE, \bibinfo{pages}{499--504}.
\newblock


\bibitem[Pu et~al\mbox{.}(2011)]%
        {pu2011user}
\bibfield{author}{\bibinfo{person}{Pearl Pu}, \bibinfo{person}{Li Chen}, {and} \bibinfo{person}{Rong Hu}.} \bibinfo{year}{2011}\natexlab{}.
\newblock \showarticletitle{A user-centric evaluation framework for recommender systems}. In \bibinfo{booktitle}{\emph{Proceedings of the fifth ACM conference on Recommender systems}}. \bibinfo{pages}{157--164}.
\newblock


\bibitem[Raglio et~al\mbox{.}(2020)]%
        {RAGLIO2020105160}
\bibfield{author}{\bibinfo{person}{Alfredo Raglio}, \bibinfo{person}{Marcello Imbriani}, \bibinfo{person}{Chiara Imbriani}, \bibinfo{person}{Paola Baiardi}, \bibinfo{person}{Sara Manzoni}, \bibinfo{person}{Marta Gianotti}, \bibinfo{person}{Mauro Castelli}, \bibinfo{person}{Leonardo Vanneschi}, \bibinfo{person}{Francisco Vico}, {and} \bibinfo{person}{Luca Manzoni}.} \bibinfo{year}{2020}\natexlab{}.
\newblock \showarticletitle{Machine learning techniques to predict the effectiveness of music therapy: A randomized controlled trial}.
\newblock \bibinfo{journal}{\emph{Computer Methods and Programs in Biomedicine}}  \bibinfo{volume}{185} (\bibinfo{year}{2020}), \bibinfo{pages}{105160}.
\newblock
\showISSN{0169-2607}
\urldef\tempurl%
\url{https://doi.org/10.1016/j.cmpb.2019.105160}
\showDOI{\tempurl}


\bibitem[Ren et~al\mbox{.}(2024)]%
        {ren2024affective}
\bibfield{author}{\bibinfo{person}{Yiren Ren}, \bibinfo{person}{Sophia~Kaltsouni Mehdizadeh}, \bibinfo{person}{Grace Leslie}, {and} \bibinfo{person}{Thackery Brown}.} \bibinfo{year}{2024}\natexlab{}.
\newblock \showarticletitle{Affective music during episodic memory recollection modulates subsequent false emotional memory traces: an fMRI study}.
\newblock \bibinfo{journal}{\emph{Cognitive, Affective, \& Behavioral Neuroscience}} \bibinfo{volume}{24}, \bibinfo{number}{5} (\bibinfo{year}{2024}), \bibinfo{pages}{912--930}.
\newblock


\bibitem[Roy et~al\mbox{.}(2008)]%
        {roy2008emotional}
\bibfield{author}{\bibinfo{person}{Mathieu Roy}, \bibinfo{person}{Isabelle Peretz}, {and} \bibinfo{person}{Pierre Rainville}.} \bibinfo{year}{2008}\natexlab{}.
\newblock \showarticletitle{Emotional valence contributes to music-induced analgesia}.
\newblock \bibinfo{journal}{\emph{Pain}} \bibinfo{volume}{134}, \bibinfo{number}{1-2} (\bibinfo{year}{2008}), \bibinfo{pages}{140--147}.
\newblock


\bibitem[Sandstrom and Russo(2010)]%
        {sandstrom2010music}
\bibfield{author}{\bibinfo{person}{Gillian~M Sandstrom} {and} \bibinfo{person}{Frank~A Russo}.} \bibinfo{year}{2010}\natexlab{}.
\newblock \showarticletitle{Music hath charms: The effects of valence and arousal on recovery following an acute stressor}.
\newblock \bibinfo{journal}{\emph{Music and Medicine}} \bibinfo{volume}{2}, \bibinfo{number}{3} (\bibinfo{year}{2010}), \bibinfo{pages}{137--143}.
\newblock


\bibitem[Schweizer et~al\mbox{.}(2020)]%
        {schweizer2020role}
\bibfield{author}{\bibinfo{person}{Susanne Schweizer}, \bibinfo{person}{Ian~H Gotlib}, {and} \bibinfo{person}{Sarah-Jayne Blakemore}.} \bibinfo{year}{2020}\natexlab{}.
\newblock \showarticletitle{The role of affective control in emotion regulation during adolescence.}
\newblock \bibinfo{journal}{\emph{Emotion}} \bibinfo{volume}{20}, \bibinfo{number}{1} (\bibinfo{year}{2020}), \bibinfo{pages}{80}.
\newblock


\bibitem[Stuckey and Nobel(2010)]%
        {stuckey2010connection}
\bibfield{author}{\bibinfo{person}{Heather~L Stuckey} {and} \bibinfo{person}{Jeremy Nobel}.} \bibinfo{year}{2010}\natexlab{}.
\newblock \showarticletitle{The connection between art, healing, and public health: A review of current literature}.
\newblock \bibinfo{journal}{\emph{American journal of public health}} \bibinfo{volume}{100}, \bibinfo{number}{2} (\bibinfo{year}{2010}), \bibinfo{pages}{254--263}.
\newblock


\bibitem[Tan et~al\mbox{.}(2024)]%
        {tan2024contrastive}
\bibfield{author}{\bibinfo{person}{J. Tan} {et~al\mbox{.}}} \bibinfo{year}{2024}\natexlab{}.
\newblock \showarticletitle{Contrastive Learning Is Spectral Clustering On Similarity Graph}. In \bibinfo{booktitle}{\emph{Proceedings of CVPR 2024}}.
\newblock


\bibitem[Thaut(2013)]%
        {thaut2013rhythm}
\bibfield{author}{\bibinfo{person}{Michael Thaut}.} \bibinfo{year}{2013}\natexlab{}.
\newblock \bibinfo{booktitle}{\emph{Rhythm, music, and the brain: Scientific foundations and clinical applications}}.
\newblock \bibinfo{publisher}{Routledge}.
\newblock


\bibitem[Vik et~al\mbox{.}(2019)]%
        {vik2019neuroplastic}
\bibfield{author}{\bibinfo{person}{Berit Marie~Dykesteen Vik}, \bibinfo{person}{Geir~Olve Skeie}, {and} \bibinfo{person}{Karsten Specht}.} \bibinfo{year}{2019}\natexlab{}.
\newblock \showarticletitle{Neuroplastic effects in patients with traumatic brain injury after music-supported therapy}.
\newblock \bibinfo{journal}{\emph{Frontiers in human neuroscience}}  \bibinfo{volume}{13} (\bibinfo{year}{2019}), \bibinfo{pages}{177}.
\newblock


\bibitem[Watson et~al\mbox{.}(1988)]%
        {watson1988development}
\bibfield{author}{\bibinfo{person}{David Watson}, \bibinfo{person}{Lee~Anna Clark}, {and} \bibinfo{person}{Auke Tellegen}.} \bibinfo{year}{1988}\natexlab{}.
\newblock \showarticletitle{Development and validation of brief measures of positive and negative affect: the {PANAS} scales}.
\newblock \bibinfo{journal}{\emph{Journal of personality and social psychology}} \bibinfo{volume}{54}, \bibinfo{number}{6} (\bibinfo{year}{1988}), \bibinfo{pages}{1063}.
\newblock


\bibitem[Wheeler et~al\mbox{.}(2011)]%
        {wheeler2011musically}
\bibfield{author}{\bibinfo{person}{Barbara~L Wheeler}, \bibinfo{person}{Estate Sokhadze}, \bibinfo{person}{Joshua Baruth}, \bibinfo{person}{Gene~Ann Behrens}, {and} \bibinfo{person}{Carla~F Quinn}.} \bibinfo{year}{2011}\natexlab{}.
\newblock \showarticletitle{Musically induced emotions: Subjective measures of arousal and valence.}
\newblock \bibinfo{journal}{\emph{Music and Medicine}} (\bibinfo{year}{2011}).
\newblock


\bibitem[Yilma et~al\mbox{.}(2024)]%
        {yilma2024artful}
\bibfield{author}{\bibinfo{person}{Bereket~A. Yilma}, \bibinfo{person}{Chan~Mi Kim}, \bibinfo{person}{Gerald~C. Cupchik}, {and} \bibinfo{person}{Luis~A. Leiva}.} \bibinfo{year}{2024}\natexlab{}.
\newblock \showarticletitle{Artful Path to Healing: Using Machine Learning for Visual Art Recommendation to Prevent and Reduce Post-Intensive Care Syndrome (PICS)}. In \bibinfo{booktitle}{\emph{Proceedings of the CHI Conference on Human Factors in Computing Systems}} (Honolulu, HI, USA) \emph{(\bibinfo{series}{CHI '24})}. \bibinfo{publisher}{Association for Computing Machinery}, \bibinfo{address}{New York, NY, USA}, Article \bibinfo{articleno}{447}, \bibinfo{numpages}{19}~pages.
\newblock
\showISBNx{9798400703300}
\urldef\tempurl%
\url{https://doi.org/10.1145/3613904.3642636}
\showDOI{\tempurl}


\bibitem[Yilma et~al\mbox{.}(2025)]%
        {yilma2025ai}
\bibfield{author}{\bibinfo{person}{Bereket~A Yilma}, \bibinfo{person}{Chan~Mi Kim}, \bibinfo{person}{Geke Ludden}, \bibinfo{person}{Thomas van Rompay}, {and} \bibinfo{person}{Luis~A Leiva}.} \bibinfo{year}{2025}\natexlab{}.
\newblock \showarticletitle{The AI-Therapist Duo: Exploring the Potential of Human-AI Collaboration in Personalized Art Therapy for PICS Intervention}.
\newblock \bibinfo{journal}{\emph{arXiv preprint arXiv:2502.09757}} (\bibinfo{year}{2025}).
\newblock


\bibitem[Yousefian~Jazi et~al\mbox{.}(2021)]%
        {yousefian2021emotion}
\bibfield{author}{\bibinfo{person}{Saba Yousefian~Jazi}, \bibinfo{person}{Marjan Kaedi}, {and} \bibinfo{person}{Afsaneh Fatemi}.} \bibinfo{year}{2021}\natexlab{}.
\newblock \showarticletitle{An emotion-aware music recommender system: bridging the user’s interaction and music recommendation}.
\newblock \bibinfo{journal}{\emph{Multimedia Tools and Applications}}  \bibinfo{volume}{80} (\bibinfo{year}{2021}), \bibinfo{pages}{13559--13574}.
\newblock


\bibitem[Zang et~al\mbox{.}(2022)]%
        {zang2022survey}
\bibfield{author}{\bibinfo{person}{Tianzi Zang}, \bibinfo{person}{Yanmin Zhu}, \bibinfo{person}{Haobing Liu}, \bibinfo{person}{Ruohan Zhang}, {and} \bibinfo{person}{Jiadi Yu}.} \bibinfo{year}{2022}\natexlab{}.
\newblock \showarticletitle{A survey on cross-domain recommendation: taxonomies, methods, and future directions}.
\newblock \bibinfo{journal}{\emph{ACM Transactions on Information Systems}} \bibinfo{volume}{41}, \bibinfo{number}{2} (\bibinfo{year}{2022}), \bibinfo{pages}{1--39}.
\newblock


\bibitem[Zhang et~al\mbox{.}(2021)]%
        {zhang2021deep}
\bibfield{author}{\bibinfo{person}{Qian Zhang}, \bibinfo{person}{Wenhui Liao}, \bibinfo{person}{Guangquan Zhang}, \bibinfo{person}{Bo Yuan}, {and} \bibinfo{person}{Jie Lu}.} \bibinfo{year}{2021}\natexlab{}.
\newblock \showarticletitle{A deep dual adversarial network for cross-domain recommendation}.
\newblock \bibinfo{journal}{\emph{IEEE Transactions on Knowledge and Data Engineering}} \bibinfo{volume}{35}, \bibinfo{number}{4} (\bibinfo{year}{2021}), \bibinfo{pages}{3266--3278}.
\newblock


\bibitem[Zhu et~al\mbox{.}(2021)]%
        {zhu2021cross}
\bibfield{author}{\bibinfo{person}{Feng Zhu}, \bibinfo{person}{Yan Wang}, \bibinfo{person}{Chaochao Chen}, \bibinfo{person}{Jun Zhou}, \bibinfo{person}{Longfei Li}, {and} \bibinfo{person}{Guanfeng Liu}.} \bibinfo{year}{2021}\natexlab{}.
\newblock \showarticletitle{Cross-domain recommendation: challenges, progress, and prospects}.
\newblock \bibinfo{journal}{\emph{arXiv preprint arXiv:2103.01696}} (\bibinfo{year}{2021}).
\newblock


\end{thebibliography}

\end{document}